# A Randomized Controlled Trial and Pilot of *Scout*: an LLM-Based EHR Search and Synthesis Platform

Michael Gao, PhD[1]; Suresh Balu, MS, MBA[1]; William Knechtle, MBA[1]; Kartik Pejavara, BS[1]; William Jeck, MD, PhD[2]; Matt Ellis, MD[2]; Jason Thieling, MD[2]; Blake Cameron, MD[2]; Jason Tatreau, MD[2]; Tareq Aljurf, MD[1,2]; Henry Foote, MD[1,2]; Michael Revoir, BS[1]; Marshall Nichols, MS[1]; Matthew Gardner, BS[1]; William Ratliff, MBA[1]; Bradley Hintze, PhD[1]; Angelo Milazzo, MD[2]; Sreekanth Vemulapalli, MD[2]

**Affiliations:**

[1]Duke Institute for Health Innovation, Duke University School of Medicine, Durham, NC, USA

[2]Duke Health, Duke University, Durham, NC, USA

**Corresponding Author:** Michael Gao, michael.gao@duke.edu

# Abstract

Clinical documentation and data retrieval within Electronic Health Records (EHRs) contribute substantially to clinician workload and burnout. To address this, we developed Scout, an LLM-based EHR search and synthesis platform that enables clinicians to query EHR data using natural language. Each response includes citations linking each claim to the original data source, facilitating easy verification of generated content. We conducted a prospective randomized, evaluator-blinded crossover trial across seven clinical specialties (20 participants, 200 structured cases). Participants completed realistic clinical tasks using either Scout or the EHR alone, with outcomes including time to completion, NASA Task Load Index workload scores, and blinded expert adjudication of accuracy, completeness, and relevance. Scout reduced task completion time by 37.6% and significantly decreased perceived workload, with the largest reductions in mental demand, effort, and temporal demand. Non-inferiority analyses showed that tasks completed with Scout maintained accuracy, completeness, and relevance relative to tasks completed with the EHR-only. A concurrent pilot deployment across over 200 users and more than 20 specialties generated over 6,600 interactions in three months, revealing diverse clinical and administrative use cases. Automated evaluation using an LLM-as-judge framework identified errors at low rates. Subsequent manual review of a subset of outputs revealed that most claims flagged by the automated judge as errors were in fact supported by the patient chart, demonstrating the importance of human validation. These findings provide early trial-based evidence that LLM-powered EHR tools can meaningfully reduce clinical and administrative workloads while maintaining output quality.

# Introduction

The burden of clinical documentation and interfacing with Electronic Health Record (EHR) systems is a primary driver of burnout for physicians and other healthcare professionals[1–16], causing significant losses in productivity and diminished patient care[17,18]. The ever-expanding volume of data stored within Electronic Health Records, duplicative information, difficulty in usability and interoperability, and policy-related mandates have all been cited as key factors contributing to the documentation burden [3,19,20]. The use of Artificial Intelligence (AI) tools based on Large Language Models (LLMs) has been shown to be a promising avenue towards alleviating some of this burden. Tools such as AI-powered ambient scribes have been shown to significantly reduce the documentation burden[21–25]. In addition, AI-powered chatbots have been shown to take this one step further by providing an LLM interface with the EHR, enabling the rapid search and synthesis of data[26–28].

However, deploying LLMs for this purpose introduces two practical challenges. First, LLM outputs can contain hallucinations or unsupported claims, which necessitates careful grounding and verification[29–33]. Second, despite the increasing adoption of LLM tools for EHRs, randomized trial evidence that these tools significantly reduce the temporal and mental burden remains limited. As these systems shift work for clinical staff from manual data curation and synthesis to verification, care is needed both to design systems that facilitate this verification and to provide rigorous evaluation.

In this work, we introduce Scout, an LLM-based platform and tool designed from the ground up at a large academic health system to provide accurate and comprehensive answers to clinical and administrative queries by searching across a patient's entire medical record. Central to the design of this platform is a focus on verification of information: each claim is accompanied by citations linking directly to the originating record content to facilitate efficient review and mitigate errors and hallucinations.

To begin characterizing Scout's impact, we pursued three aims. First, we evaluated whether Scout helps reduce the time and perceived cognitive burden required to complete realistic clinical and administrative tasks. Second, we assessed whether these efficiency gains can be achieved without compromising the accuracy, completeness, or relevance of task outputs. Third, we characterized usage patterns, user needs, and monitoring strategies that emerge as these tools transition from evaluation to deployment to inform their safe and sustainable integration into clinical practice. To address these aims, we conducted a randomized crossover trial across multiple specialties and report findings from a pilot deployment at our institution.

# Methods

## Study Overview

We conducted a prospective, randomized, evaluator-blinded, two-period crossover trial at a large academic health system to evaluate the utility of Scout compared to usual workflows using the EHR alone. Structured tasks were created by expert physicians and completed by participants that included faculty physicians, post-graduate trainees, advanced practice providers (APPs), and other clinical staff. We assessed the time it took to complete these tasks and self-reported burden using the NASA Task Load Index (NASA-TLX)[34] instrument. In addition, completed structured outputs were then reviewed by senior physicians in a blinded manner to assess for overall quality. All aspects of the trial were approved under the Duke University Health System Institutional Review Board (DUHS IRB) in accordance with Protocol #Pro00118622.

## Scout

Scout is an AI-driven platform and web application developed at Duke Health that facilitates the efficient search and synthesis of EHR data while providing full data provenance using citations. The Scout application searches and analyzes data sources such as unstructured text (*e.g.* notes, pathology reports, diagnostic reports, etc.), structured data (*e.g.* laboratory values, vitals signs), and uploaded external documents (*e.g.* PDFs). These data are then synthesized to create an output which addresses the user's query. This is done by a combination of agentic search and task planning, which is handled by the main LLM orchestrator. Once relevant information and citations are collected, this information is used to generate the answer in a user-specified format (*e.g.* formatted summary, note template). Each claim made by Scout is accompanied by a citation that links to the original data used to generate the claim. For all notes that are referenced, the relevant span of text which contains the information used during answer generation provided in the form of an *evidence card*. Providing this transparency has a two-fold benefit: it (1) facilitates easy verification of source information and (2) reduces potential errors and hallucinations by grounding each claim to the original source data. The orchestrator routes queries to fast-paths and slow-paths depending on the complexity of the task and patient at hand, allowing fast responses for simple queries and supporting extended processing for complex tasks. Figure 1 illustrates Scout's process and how data flows through the application. Additional technical details can be found in Appendix A.

## Study Design

To evaluate the utility and performance of Scout, a team of seven senior physicians representing seven different specialties (Nephrology, Transplant Surgery, Oncology, Adult Cardiology, Pediatric Cardiology, Emergency Medicine, Psychiatry) created structured tasks intended to mimic realistic clinical/administrative workflows. For each task, the supervising physician selected 10 patient charts with which they were familiar for participants to complete the task against. Completed outputs were scored by blinded evaluators using a standardized rubric (see Blinded Quality Adjudication). An example task is shown in Figure 2. Task descriptions, participant instructions, surveys, and the scoring rubric are provided in Appendices B–E. Participants were then selected from clinical team members who routinely complete similar tasks in their day-to-day work. For each use case, participants were chosen with the aim of reducing variations in both role and experience level wherever feasible. At least two participants were enrolled for each use case. Potential participants with prior knowledge of the selected patients were excluded from participating as determined via self-assessment.

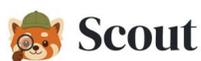 Scout

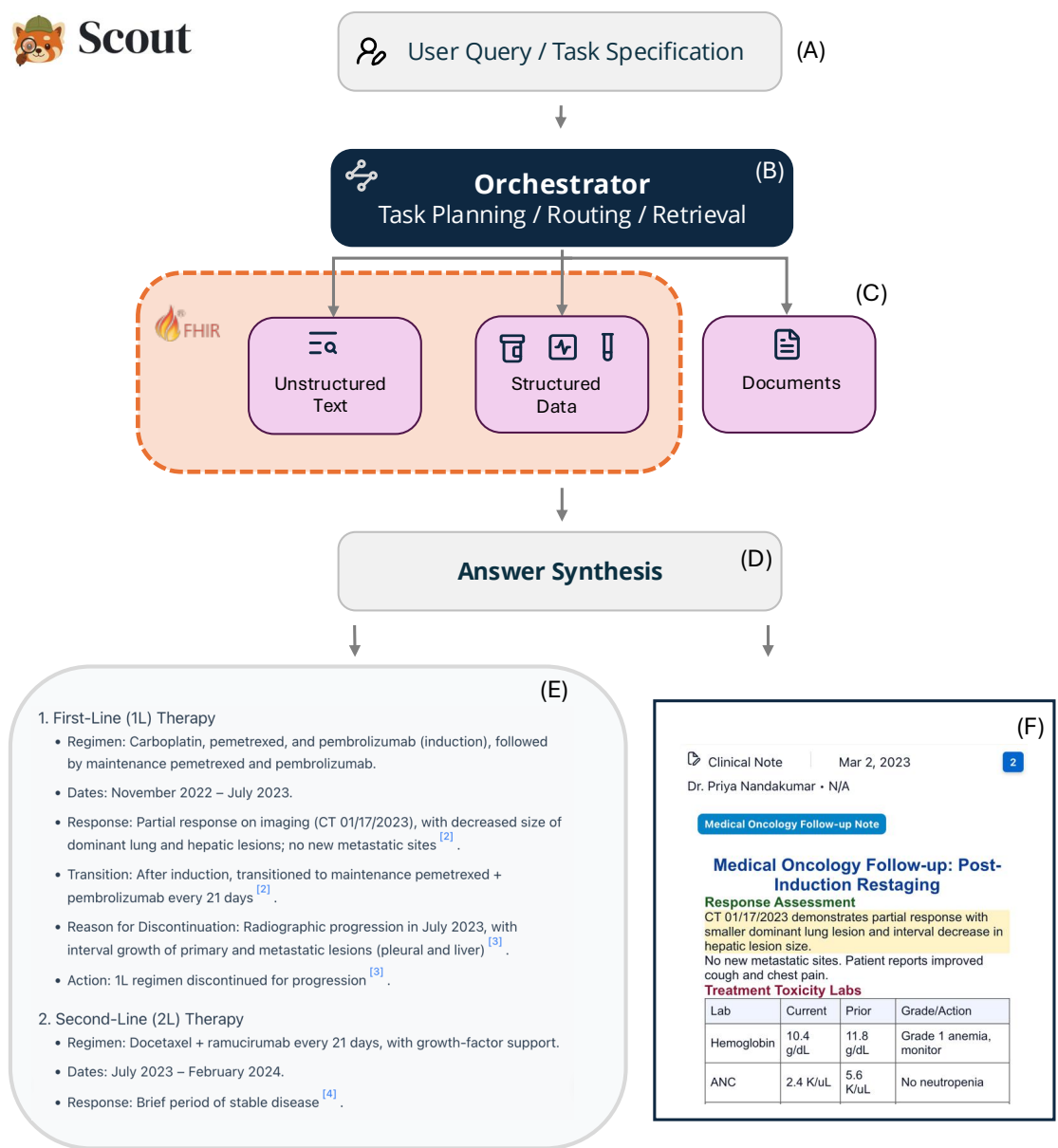

**Figure 1**: **Scout Data Flow Architecture.**
**(A)** The user specifies a prompt in free-text form, specifying what task Scout should complete. **(B)** The LLM-based orchestrator takes the prompt, evaluates which data sources and tools need to be invoked (such as subagents to identify relevant labs, medications, relevant spans from notes, etc.) and creates a plan for execution. **(C)** Relevant data sources (FHIR-based resources) and user-uploaded content is returned to the main LLM agent. **(D)** After all relevant information has been included, the answer is generated with appropriate citations and guardrails (including safety checks). **(E)** The answer, complete with citations is displayed to the user. **(F)** *Evidence Cards*, views into the original data cited by Scout, are included to facilitate verification of information.

.

The study employed a two-period randomized crossover trial. Within each use case, participants were randomized with an allocation of 1:1 to one of two sequences to mitigate learning and time-related effects: (1) Scout for the first block of five cases followed by EHR-only for the second block of five cases or (2) EHR-only for the first block followed by Scout for the second block. During the Scout block, participants were asked to verify their work either in the Scout application or with the help of the EHR. The EHR-only cases were required to be completed without the use of any additional software. Participants were given standardized instructions regarding task completion and time capture (Appendix C). Prior to the commencement of the trial, participants received a brief training session, which included a training video and a 30-minute meeting to outline the study and address any questions regarding its completion.

Upon task completion, all structured outputs were adjudicated by a senior physician evaluator using a standardized rubric applied uniformly across use cases, in which accuracy, completeness, and relevance were each scored on an 11-point scale (0–10); the full rubric is provided in Appendix E. For each use case, the evaluating physician was the same physician who authored the cases for that use case. This design was chosen to ensure that outputs were assessed by clinicians with direct domain expertise in the specific workflow and familiarity with the expected clinical content for each patient, which we judged essential for reliable scoring of open-ended outputs.

For the following patients, please fill out the following template in order to prepare for a tumor board discussion

- Date of initial presentation for current oncologic disease
- Tumor type, including histologic subtype and any relevant pathologic grade
- Tumor primary site
- Sites of active disease
- Clinical stage
- Pathologic stage
- Most recent tumor directed therapy, including immunotherapy, cytotoxic chemotherapy,
- Targeted therapy, and radiotherapy
- Prior therapy history, including immunotherapy, cytotoxic chemotherapy, targeted therapy and radiotherapy, with start and end dates
- Previous oncologic surgeries or biopsies
- History of involvement in clinical trials, including the name of the study drugs, where relevant.

**Figure 2: Example Task Related to Tumor Board Discussions.** Each participant within the tumor board use case would fill out the following information. Participants were randomized to complete one block of five cases with Scout and the other block with EHR-only.

To preserve evaluator blinding, all outputs were stripped of participant identifiers and arm assignments prior to adjudication and were presented to evaluators in randomized order. In addition, outputs were conformed to a standardized task template to prevent formatting differences from revealing the study arm; where a participant's submission deviated from the template, it was reformatted prior to adjudication to match the standard structure. Evaluators were blinded to both participant identity and study arm throughout the scoring process.

## Endpoints

Our hypothesis was that AI tools such as Scout would reduce the amount of time taken to complete tasks involving the search and synthesis of clinical data, reduce the overall burden on clinical staff, and simultaneously maintain the quality of the task outputs. Our primary endpoints were the time to complete structured patient cases and the workload burden measured using raw NASA-TLX score after excluding physical demand. Secondary outcomes are mental demand, temporal demand, performance, effort and frustration as assessed by NASA-TLX subscale scores. To ensure that the AI tools would not harm the quality of the responses, we additionally report non-inferiority endpoints for Accuracy, Completeness, and Relevance as judged by expert evaluators as secondary outcomes. In addition, we are interested in patterns that emerge when users apply these tools in their daily work. We report pilot metrics, including types of users and usage patterns.

## Statistical Analysis

### Sample Size

Sample size calculations were done using simulation-based power analyses designed to match the planned primary analysis model. We conducted Monte Carlo simulations using the same model specification as the primary analysis: a linear mixed-effects model with Tool as the fixed effect of interest, patient case (MRN) as a fixed effect to control for case-level difficulty, and a random intercept for participant.

For the primary time endpoint, pilot observations indicated baseline completion times of approximately 5–15 minutes across participants and cases, with Scout estimated to reduce time by roughly half. We conservatively targeted a 25% reduction in geometric mean time (ratio of geometric means = 0.75; $\Delta \approx$ -0.29 on the log scale). Because pilot data were limited and insufficient to formally estimate variance components, we evaluated power across a grid of plausible values for the between-participant SD ($\sigma_u \in \{0.20, 0.30, 0.40\}$ on the log scale) and residual SD ($\sigma_e \in \{0.20, 0.30, 0.35\}$ on the log scale), informed by the observed range of pilot completion times and expected within-person variability.

For the co-primary NASA-TLX endpoint, we targeted a reduction of 8 points on the raw score (0–100 scale), based on published benchmarks from studies of AI documentation tools in clinical settings. Variance assumptions ($\sigma_u \in \{5, 8, 12\}$; $\sigma_e \in \{10, 14, 18\}$ points) were informed by reported effect sizes and variability in comparable studies[35,36].

For each scenario, 5,000 datasets were simulated under the planned model, randomizing participants to Scout-first or Control-first sequences. The analysis model was fit to each simulated dataset, and power was estimated as the proportion of simulations in which the two-sided p-value for the Tool effect was below 0.05. Across all variance scenarios, power for the time endpoint exceeded 0.99, and power for the NASA-TLX endpoint exceeded 0.85 (worst case: 0.852 at $\sigma_e = 18$). With 20 participants across 7 use cases, the study met or exceeded 80% power for both primary endpoints under all considered scenarios. Full simulation details and code are provided in Appendix F.

**Primary Analysis**
Within each use case, the same 10 patient cases were presented in fixed order to all participants, with cases 1–5 forming Block 1 and cases 6–10 forming Block 2. Participants were randomized to complete Block 1 with Scout and Block 2 with EHR-only, or vice versa. Because case order was fixed across participants, we modeled each patient case (MRN) as a fixed effect to account for case-specific difficulty, which also absorbs any between-block differences in case complexity.

For all endpoints, we fit linear mixed-effects models with fixed effects for Tool (Scout vs. Control) and patient case (MRN), and a random intercept for participant. Each model was fit at the case level with one observation per participant-case. Time-to-completion was log-transformed, and results are reported as the ratio of geometric means (Scout vs. Control). This model structure was applied to all endpoints, including NASA-TLX outcomes (raw score and prespecified subscales) and evaluator-assigned scores for accuracy, completeness, and relevance. As a prespecified sensitivity analysis, we fit an alternative model replacing MRN fixed effects with a period indicator (Block 1 vs. Block 2) and a Tool × Period interaction to assess potential carryover effects.

**Non-Inferiority Testing**
For quality outcomes, we conducted prespecified non-inferiority analyses using a margin of −0.5 point for the mean difference (Scout – Control). Each rubric uses an 11-point scale (0–10) in which each integer corresponds to a qualitatively distinct tier of clinical usability. For example, on the accuracy rubric, a score of 8 ("Very Good") denotes no clinically significant errors, whereas a score of 7 ("Good") introduces an error of moderate clinical relevance. A margin of −0.5 point represents half the distance between adjacent categories, ensuring that any acceptable degradation would not on average shift outputs into a lower tier. In preliminary testing, scores for both conditions generally fell in the 6–10 range, making the effective operating range approximately 5 points and the chosen margin roughly 10% of that range. This threshold was further endorsed through consensus among the evaluating physicians as below the level that would be perceptible in practice. Non-inferiority was concluded if the lower bound of the two-sided 95% confidence interval for (Scout – Control) exceeded −0.5. Degrees of freedom and 95% confidence intervals were computed via Satterthwaite approximations.

**Multiple Comparisons**

We controlled the family-wise error rate (two-sided $\alpha$=0.05) separately within each of three outcome families, reflecting their different inferential goals: primary outcomes (time to completion, NASA-TLX raw score; 2 total outcomes), NASA-TLX secondary outcomes (5 component outcomes), and NI outcomes (accuracy, completeness, relevance; 3 outcomes) using the Holm-Bonferroni procedure. All statistical analyses including randomization scheme were conducted using R (v. 4.1.3). A full list of packages and versions used in the statistical analysis is provided in Appendix G.

## Automated Quality Monitoring

In addition to the human adjudications, we evaluated Scout's outputs across all 70 tasks using an LLM-as-judge[37] approach based on a modified version of the *VeriFact*[38] framework, adopting exact prompts from analyses of similar tools[26]. This automated evaluation proceeded in three stages. First, each output was decomposed into individual claims. Second, each claim was compared against the patient's chart to determine whether it was supported by the reference text. Third, unsupported claims were classified as either hallucinations (claims lacking support in the source material) or inaccuracies (claims directly contradicting the source text). OpenAI's GPT-4.1 served as the judge model (chosen for its balance of performance and large context window), evaluating all claims using standardized, use-case-specific prompts. We report the frequency of hallucinations and inaccuracies across all tasks. To assess the reliability of this LLM-as-judge approach, we also manually validated every claim for a subset of 5 patients and present those findings. Prompts used in this framework can be found in Appendix H.

## Pilot Phase

To characterize usage patterns and inform the safe integration of Scout into clinical practice, we also deployed the tool to a convenience sample of pilot users. Enrollment began in September 2025, and data were collected through the end of the year. We recorded users' clinical titles, specialties or service lines, and number of tasks submitted. We also mapped the types of clinical questions asked and the feedback provided within the application. All users completed onboarding consisting of a video overview of the tool and its intended use, and signed IRB-approved consent forms prior to access. Dedicated feedback channels were established, including email, a Microsoft Teams channel, and in-app feedback.

## Results

In total, 20 participants were recruited across 7 service lines (Nephrology, Transplant Surgery, Oncology, Adult Cardiology, Pediatric Cardiology, Emergency Medicine, Psychiatry). 7 attending physicians, 3 post-graduate trainees, 3 advanced practice providers, and 7 nurses or specialized clinical staff were recruited to the study. Out of 200 possible cases (10 per participant), 189 were completed entirely successfully, with failure to complete the appropriate surveys accounting for 1 missing timing and 11 missing NASA-TLX submissions. For each outcome, we fit models including only observations with non-missing data for that outcome and its required covariates (complete-case analysis per outcome). All 200 structured outputs were received for non-inferiority outcomes. The randomization process is summarized in Figure 3.

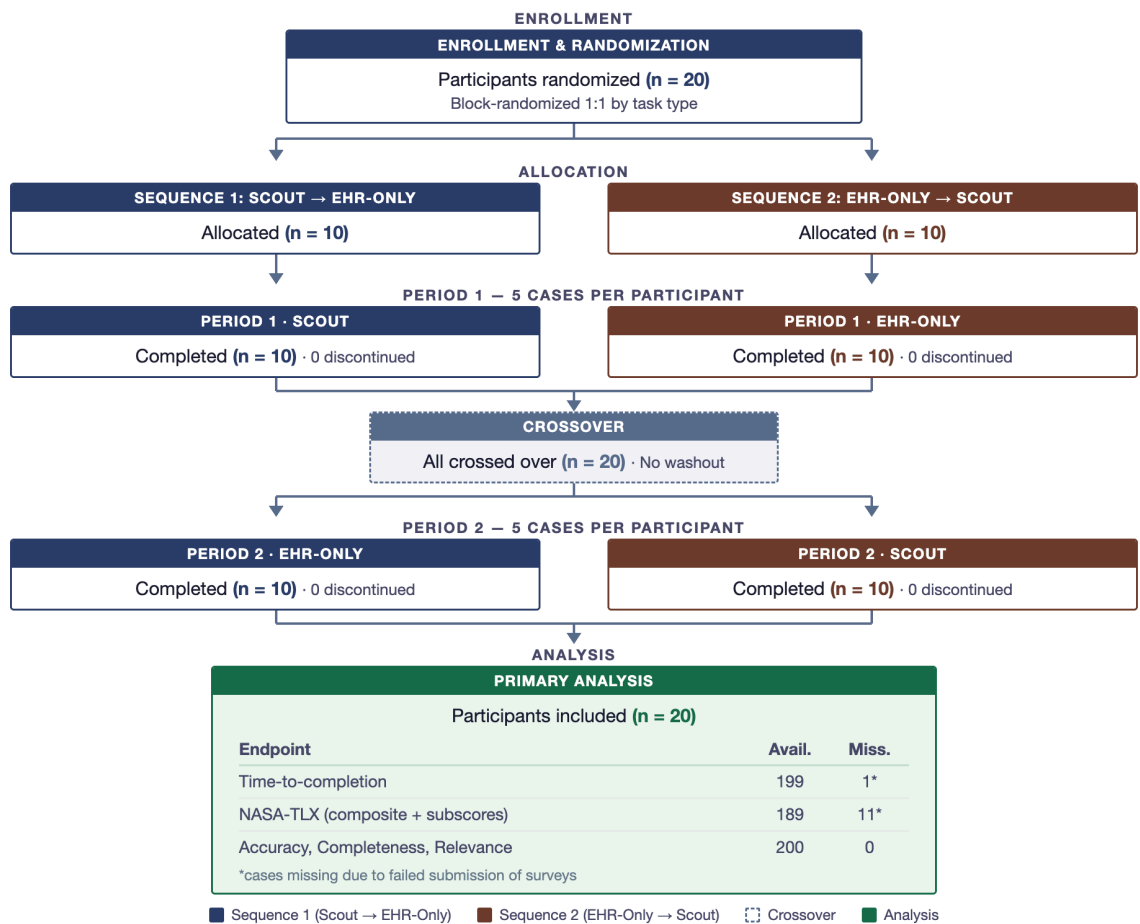

**Figure 3: Trial Participation and Randomization.** Each participant within a use case was allocated the same 10 patients (randomized once at the start of each use case and consistent across participants within use case). Within use cases, participants were randomized to sequence 1 (Scout for first 5 cases) or sequence 2 (EHR-only for the first 5 cases). After completing the first block of 5 cases, participants crossed over into the other arm.

Table 1 reports the full breakdown of roles and use cases after the randomization process was complete.

| Use Case/Specialty | Faculty Physicians | Post-graduate Trainees | APP (PA + NP) | Other Clinical Staff | Total |
|---|---|---|---|---|---|
| Behavioral Health | 2 | 0 | 0 | 0 | 2 |
| Adult Cardiology (Chest Pain) | 0 | 1 | 1 | 0 | 2 |
| Emergency Medicine | 2 | 0 | 2 | 0 | 4 |
| Oncology (Molecular Tumor Board) | 0 | 1 | 0 | 1 | 2 |
| Nephrology | 2 | 0 | 0 | 0 | 2 |
| Pediatric Cardiology | 1 | 1 | 0 | 0 | 2 |
| Transplant | 0 | 0 | 0 | 6 | 6 |

**Table 1: Breakdown of Roles for Participants Within Use Case**. Care was taken to ensure that participants within the same use case had the same level of experience with completing the task or similar tasks. Other clinical staff included transplant coordinators and clinical researchers.

## Task Completion Time and Perceived Workload

### Time Endpoint
Across all use cases, Scout was associated with significantly faster task completion. The model-based geometric mean time to completion was 9.67 minutes (95% CI, 8.29 to 11.28) under usual EHR workflow and 6.03 minutes (95% CI, 5.17 to 7.03) with Scout. This corresponded to a ratio of geometric means (Scout/Control) of 0.624 (95% CI, 0.565 to 0.689, p < 0.001), indicating that completion times were 37.6% shorter with Scout than with EHR-only across workflows. Figure 4 illustrates the raw differences in time saved for each use case. The sensitivity analysis evaluating potential carryover effects using a period indicator and a Tool×Period interaction provided no evidence that Scout's effect differed between blocks (all interaction p-values >0.20; see Appendix I).

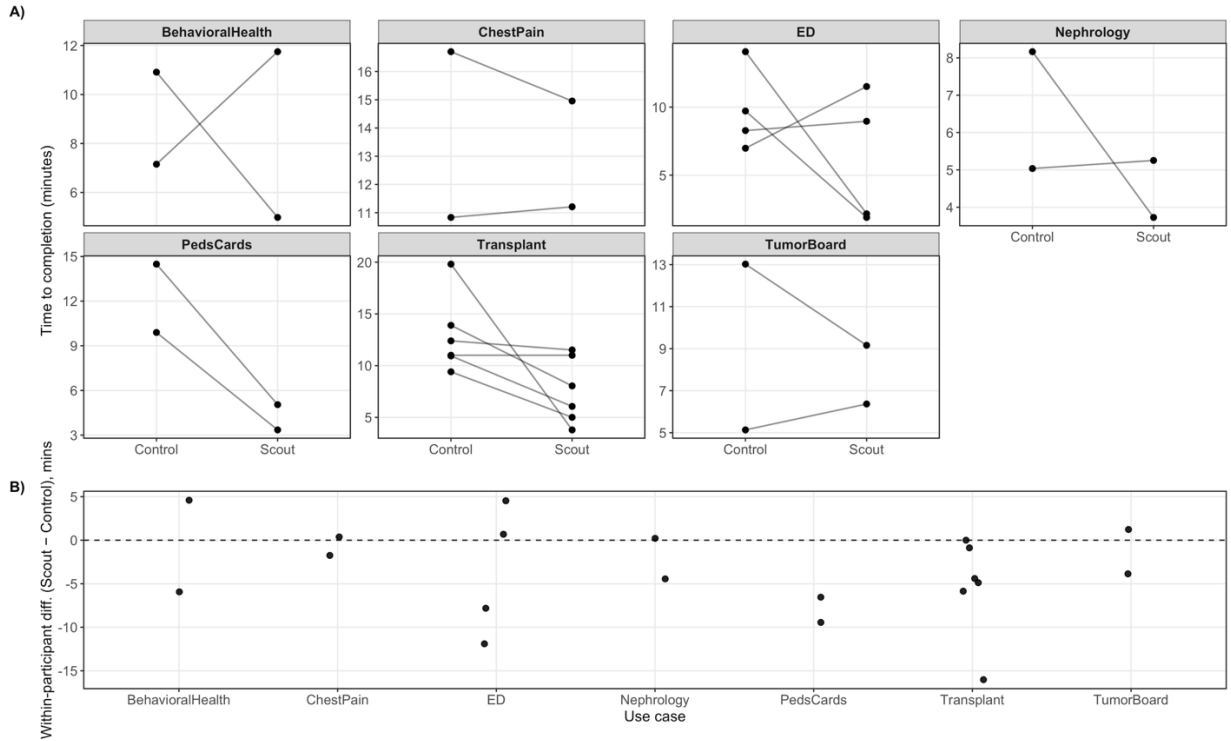

**Figure 4: Absolute Within-Participant Time Differences by Use Case.** Panel A shows paired completion times (minutes) for each participant under EHR-only workflow (Control) and Scout, with points connected within participant. Panel B shows the corresponding within-participant difference (Scout − Control), where values below 0 indicate faster completion with Scout. The trial was powered to detect an overall effect of Scout across use cases; per-use-case sample sizes are small (n = 2–6 participants each), so this breakdown is descriptive and should not be used to draw conclusions about individual use cases. Times include prompting the system and awaiting LLM responses. Detailed instructions for how time to completion was measured are provided in Appendix C.

## Workload Endpoints

Scout was associated with significantly lower perceived workload. The adjusted mean NASA-TLX raw score was 32.81 (95% CI, 26.17 to 39.46) for EHR-only and 24.07 (95% CI, 17.41 to 30.73) for Scout, yielding a mean difference of −8.74 points (95% CI, −12.02 to −5.46, adjusted p < 0.001).

Among prespecified NASA-TLX subscales, Scout was associated with significantly lower effort (−18.0 points, Holm-adjusted p < 0.001), mental demand (−16.6 points, Holm-adjusted p < 0.001), and temporal demand (−11.0 points, Holm-adjusted p < 0.001). No meaningful difference was found for frustration (−3.36 points, Holm-adjusted p = 0.192). The reversed performance subscale showed a small increase of 4.80 points (Holm-adjusted p = 0.003), suggesting that participants perceived slightly lower performance when using Scout, possibly reflecting uncertainty about the verification process with a new tool. Full results are presented in Table 2.

| Outcome | Scale | EHR-Only (Control) | Scout | Effect Size | Adjusted p-value |
|---------|-------|--------------------|-------|-------------|------------------|
| NASA-TLX Raw | (0-100), lower is better | 32.81 (26.17, 39.46) | 24.07 (17.41, 30.73) | −8.74 (−12.0, −5.46) | <0.001 |
| NASA-TLX Effort | (0-100), lower is better | 41.58 (33.98, 49.18) | 23.55 (15.93, 31.16) | -18.03 (-22.70, -13.37) | <0.001 |
| NASA-TLX Mental Demand | (0-100), lower is better | 37.58 (30.82, 44.35) | 20.98 (14.20, 27.75) | -16.61 (-21.19, -12.02) | <0.001 |
| NASA-TLX Performance | (0-100), lower is better | 20.09 (14.61, 25.56) | 24.89 (19.42, 30.36) | 4.80 (1.86, 7.75) | 0.003 |
| NASA-TLX Temporal Demand | (0-100), lower is better | 33.42 (24.61, 42.22) | 22.41 (13.60, 31.21) | -11.01 (-15.30, -6.73) | <0.001 |
| NASA-TLX Frustration | (0-100) lower is better | 31.34 (22.61, 40.07) | 27.98 (19.25, 36.70) | -3.36 (-8.44, 1.72) | 0.192 |

**Table 2: NASA-TLX Raw and Subscale Scores.** NASA-TLX Raw is the unweighted average of all of the components listed in the table (excludes physical demand). Adjusted means (95% CI) are shown for each condition, along with the mean difference (Scout – Control) and Holm-adjusted p-values. Scout was associated with significantly lower mental demand, effort, and temporal demand compared to EHR-only workflow.

## Output Quality

Scout met prespecified non-inferiority criteria across all three quality domains. For each domain, the lower bound of the 95% confidence interval for the adjusted mean difference (Scout − Control) exceeded the prespecified margin of −0.5 points (Figure 6). The confidence interval for completeness also excluded zero, suggesting superiority, though this was not a prespecified test. Full model estimates are reported in Appendix J.

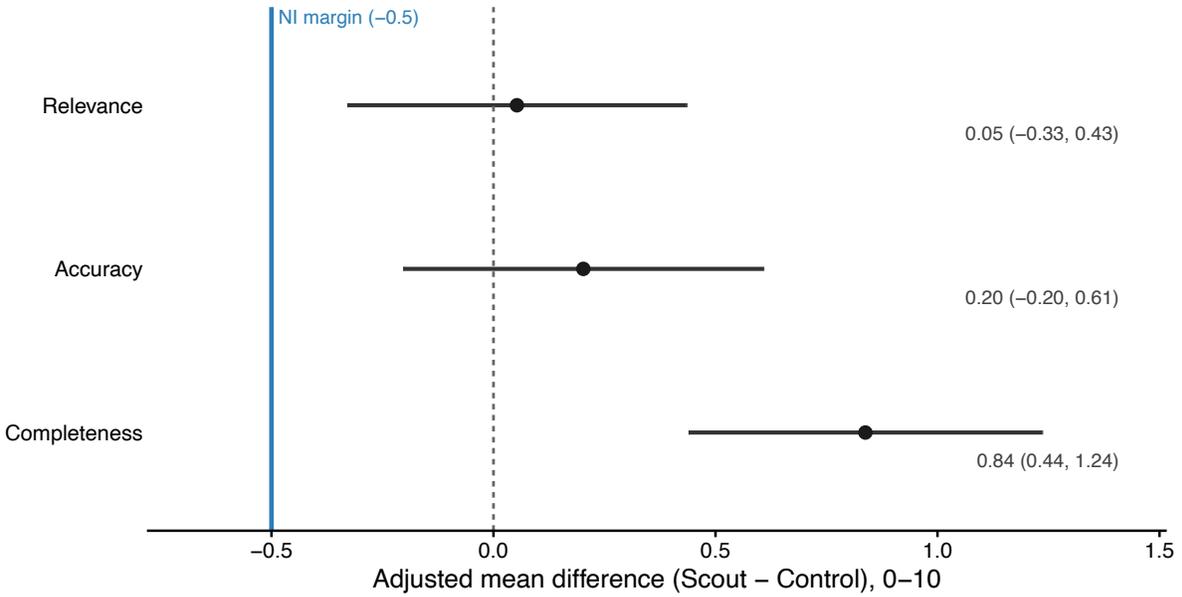

**Figure 6: Non-inferiority of Scout vs Control for quality domains.** Points show model-adjusted mean differences (Scout − Control) on the 0–10 scale from mixed-effects models adjusting for case (MRN) with a random intercept for participant; horizontal lines are 95% confidence intervals. The dashed vertical line at 0 indicates no difference (superiority boundary), and the solid vertical line at −0.5 marks the prespecified non-inferiority margin.

## Pilot Deployment

### User Adoption

From September 22, 2025, through December 31, 2025, Scout was used across an academic multi-hospital health system by 82 faculty physicians, 48 residents and post-graduate trainees, 23 nurses, and 45 other clinical staff such as pharmacists and physical therapists. Table 3 shows a breakdown of these early users by specialty and role. During the pilot period, users submitted 6641 queries to Scout. By the end of the pilot period, Scout had approximately 70 weekly active users.

| Specialty | Users | Faculty Physicians | APP | Nurse | Trainees | Other Clinical Staff |
|---|---|---|---|---|---|---|
| **Totals** | **213** | **82** | **15** | **23** | **48** | **45** |
| **Surgery & Anesthesia** | 23 | 5 | 1 | 12 | 2 | 3 |
| **Oncology** | 21 | 6 | 1 | 4 | 2 | 8 |
| **Psychiatry** | 31 | 14 | 3 | 2 | 7 | 5 |
| **Genetics** | 13 | 6 | 0 | 0 | 3 | 4 |
| **Radiology** | 21 | 8 | 0 | 0 | 13 | 0 |
| **Hospital Medicine** | 19 | 11 | 1 | 0 | 7 | 0 |
| **Neurology** | 6 | 2 | 2 | 0 | 2 | 0 |
| **Heart & Vascular** | 10 | 3 | 1 | 0 | 4 | 2 |
| **Pulmonary** | 8 | 3 | 0 | 1 | 0 | 4 |
| **Endocrinology** | 5 | 3 | 1 | 0 | 0 | 1 |
| **Gastroenterology** | 6 | 3 | 0 | 1 | 1 | 1 |
| **Emergency** | 5 | 2 | 2 | 0 | 1 | 0 |
| **Rheumatology** | 2 | 1 | 0 | 0 | 1 | 0 |
| **Pharmacy** | 5 | 0 | 0 | 0 | 0 | 5 |
| **Geriatric** | 3 | 2 | 1 | 0 | 0 | 0 |
| **Admin** | 11 | 1 | 2 | 1 | 0 | 7 |
| **PT/OT/ST/Audio** | 4 | 0 | 0 | 0 | 0 | 4 |
| **Nephrology** | 4 | 3 | 0 | 0 | 1 | 0 |
| **Primary Care & Pediatrics** | 6 | 4 | 0 | 0 | 2 | 0 |
| **Infectious Disease** | 4 | 3 | 0 | 1 | 0 | 0 |
| **Critical Care** | 1 | 0 | 0 | 1 | 0 | 0 |
| **Women's** | 3 | 1 | 0 | 0 | 1 | 1 |
| **Dermatology** | 2 | 1 | 0 | 0 | 1 | 0 |

**Table 3: Breakdown of Scout User Roles.** APPs include physician assistants and nurse practitioners. The nurses' category excludes nurse practitioners. 'Other Clinical Staff' includes titles such as pharmacist, physical therapist, and audiologist.

## Task Characterization

We mapped each of 6,641 user questions to one of 18 categories to characterize the clinical and operational needs driving use. Survey categories were manually developed to reflect clinical, research, and operational interests and use cases at our institution, informed by discussions with early users about the breadth of Scout's potential applications. The full list of user questions was submitted to OpenAI's GPT-5.2 model to assign each response to our manually-developed categories. Across the 6,641 questions, the most frequent requests were for broad chart review and summarization: 54% involved comprehensive patient summaries, medical histories, review of systems, or other general clinical documentation intended to support chart review. Point-of-care information retrieval and temporal pattern analysis were also common. A full breakdown of task categories is shown in Table 4. Prompts used to categorize user questions can be found in Appendix K.

| Task Focus | User Questions (n) |
|---|---|
| Clinical Summary and Multi-Chart Review | 3564 |
| Point-of-Care Information Retrieval | 2043 |
| Registries and Forms | 1110 |
| Temporal Pattern Analysis | 1375 |
| Clinical Reasoning and Decision Support | 567 |
| Consults, Internal Handoffs, and Intra-encounter coordination | 182 |
| Quality and Safety | 127 |
| Admissions and Inter-system movement | 117 |
| Discharge and Transitions of Care | 80 |
| Clinical Trials and Study-Fit | 96 |
| Preventive Care | 32 |
| | |
| **Chart Sub-Element Focused On** | **User Questions (n)** |
| Imaging | 263 |
| Medication | 371 |
| Labs or Vitals | 207 |
| Diagnoses or Problem Lists | 247 |
| Psychiatric or Behavioral | 259 |

Table 4: **Task Categories During Pilot Period**. Classification of 6,641 user queries from the pilot deployment, organized by task focus (upper section) and chart sub-element (lower section). Categories were developed from discussions with early users and assigned using GPT-5.2. Individual queries could be flagged under multiple categories, so counts are not mutually exclusive.

Overall, this distribution suggests Scout was primarily used for high-frequency cognitive tasks tied to longitudinal chart review, alongside more targeted information retrieval and documentation needs.

## User Feedback

Users provided in-app feedback on response quality using a 1–5 rating scale (higher is better). Across rated interactions (n=277), the distribution of quality ratings was: 1 (n=39), 2 (n=21), 3 (n=44), 4 (n=64), and 5 (n=109). Using a predefined positive threshold of 4–5, 62.5% of rated responses were positive (173/277). Feedback was provided for only 4.2% of total interactions (277/6,641), so these results may not be representative of overall user experience. Lower ratings often reflected the dual role of feedback as both a satisfaction signal and a mechanism for issue reporting. Overall, many early users felt that Scout provided meaningful value in their daily work. We include a collection of user feedback and anecdotes as illustration of the perceived value of Scout in improving clinical and administrative workflows in Figure 7.

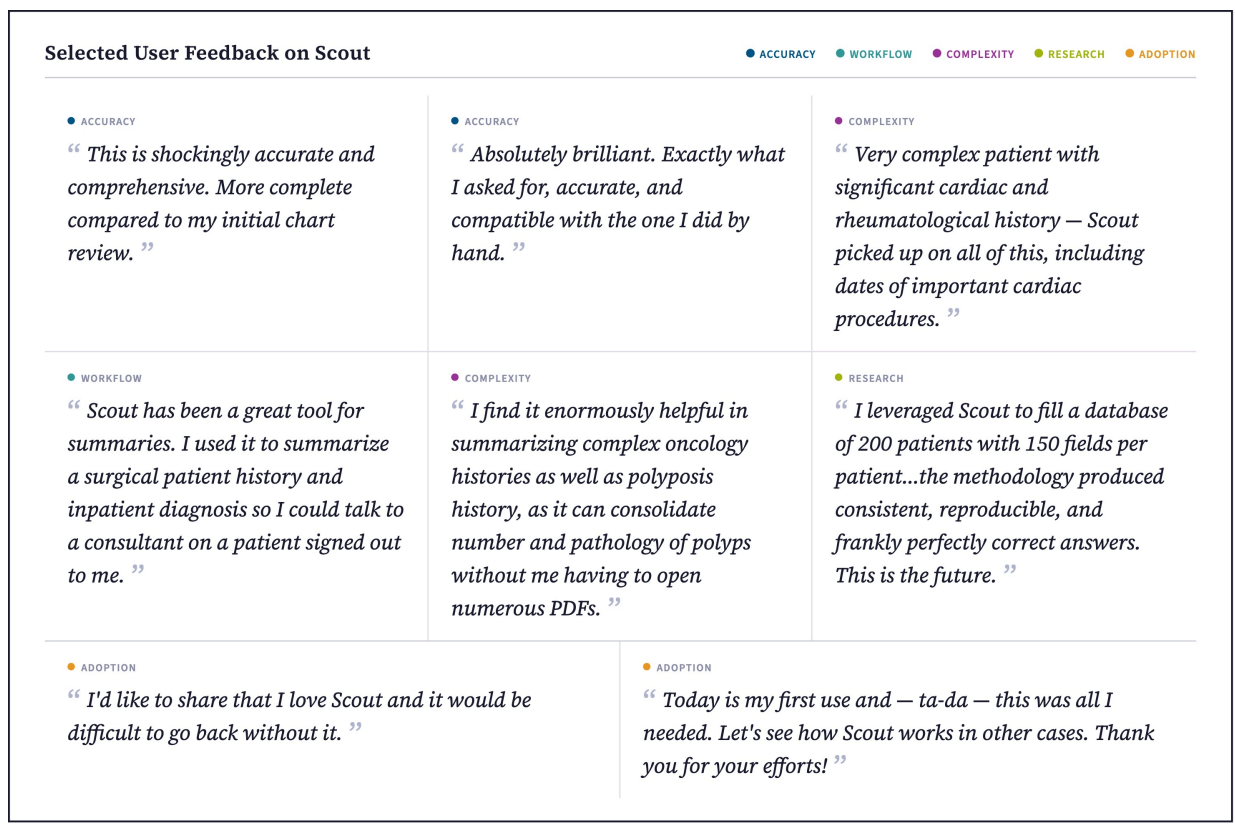

**Selected User Feedback on Scout**   ● ACCURACY ● WORKFLOW ● COMPLEXITY ● RESEARCH ● ADOPTION

● ACCURACY
*" This is shockingly accurate and comprehensive. More complete compared to my initial chart review. "*

● ACCURACY
*" Absolutely brilliant. Exactly what I asked for, accurate, and compatible with the one I did by hand. "*

● COMPLEXITY
*" Very complex patient with significant cardiac and rheumatological history — Scout picked up on all of this, including dates of important cardiac procedures. "*

● WORKFLOW
*" Scout has been a great tool for summaries. I used it to summarize a surgical patient history and inpatient diagnosis so I could talk to a consultant on a patient signed out to me. "*

● COMPLEXITY
*" I find it enormously helpful in summarizing complex oncology histories as well as polyposis history, as it can consolidate number and pathology of polyps without me having to open numerous PDFs. "*

● RESEARCH
*" I leveraged Scout to fill a database of 200 patients with 150 fields per patient...the methodology produced consistent, reproducible, and frankly perfectly correct answers. This is the future. "*

● ADOPTION
*" I'd like to share that I love Scout and it would be difficult to go back without it. "*

● ADOPTION
*" Today is my first use and — ta-da — this was all I needed. Let's see how Scout works in other cases. Thank you for your efforts! "*

**Figure 7: Qualitative User Feedback**. Users appreciated Scout's accuracy, ability to deal with complex patients, and how it integrated into their workflows. Users of Scout evaluated its ability to aid in research, clinical trials screening, and many other use cases.

## Automated Quality Monitoring

While the blinded expert evaluation serves as the primary assessment of output quality, scalable automated monitoring will be essential as Scout is deployed more broadly. The modified VeriFact[38] pipeline identified 27 hallucinations (0.38 per output) and 24 inaccuracies (0.34 per output), with 47 outputs (67%) containing neither. Further classification using prompts from prior work found that none of the identified errors would result in moderate harm, defined as adversely affecting functional ability or quality of life at a level below severe harm[26]. These results are directionally consistent with the expert evaluations and offer a preliminary signal that automated monitoring may be feasible at scale.

Upon manual inspection of a subset of these LLM-as-judge evaluations, many of these hallucinations and inaccuracies were identified as mistakes by the LLM-as-judge system itself. Across a sample of five patient outputs, the LLM-as-judge system identified 187 discrete claims generated by Scout. Of these, 170 were classified as supported and 17 as unsupported (9 hallucinations and 8 inaccuracies). Manual adjudication confirmed that all 170 claims labeled as "fully supported" were correctly classified. However, 11 of the 17 claims labeled as unsupported were in fact supported by the source text, indicating false-positive error by the LLM-as-judge system. Several patterns were observed among these misclassifications. In multiple cases, the LLM-as-judge failed to identify supporting evidence present in the patient chart despite Scout correctly citing relevant information. In other instances, the judge model interpreted Scout's synthesized phrasing as inaccurate despite the underlying clinical statement being correct. We also observed cases where the judge model flagged omissions of information that were not relevant to the prompt being answered. These findings suggest that LLM-as-judge evaluations may be sensitive to prompt design, model behavior, and interpretation of clinical language. We report representative examples of these conflicts in Appendix L.

Among the six claims that were determined to be unsupported, two primary error modes were observed: reference of outdated information when more recent documentation contained contradictory findings, and misclassification of symptoms in clinical lists. However, two of the six "unsupported" claims were factually correct but lacked appropriate source citations in Scout's output, preventing the LLM-as-judge system from identifying supporting evidence in the reference text. In total, human review found no instances in which a claim was generated entirely without any related content in the patient's chart, though in some cases the supporting text was outdated, miscategorized, or insufficiently cited. Although the LLM-as-judge system exhibited a notable false-positive rate in this small sample, it correctly identified all supported claims and flagged all the unsupported claims in our human-reviewed sample. With improvements, these types of systems can serve as practical and scalable screening tools. The reviewed outputs form the beginning of a gold-standard dataset that can be used to assess the ongoing accuracy and quality of the system as changes are introduced, such as prompt updates, addition of new workflows, or model upgrades.

Our evaluation and monitoring system, informed by this work, consists of 1) clinical expert review of new use cases consisting of manual adjudication, 2) addition of these cases to a gold-standard dataset, and 3) periodic and as-needed evaluation with LLM-as-Judge. Figure 8 illustrates this workflow, which we adopt as new use cases are deployed and require ongoing validation. This allows for the careful review of use cases coupled with the ability to spot any changes in functionality and establishes a feedback loop that helps identify and correct errors.

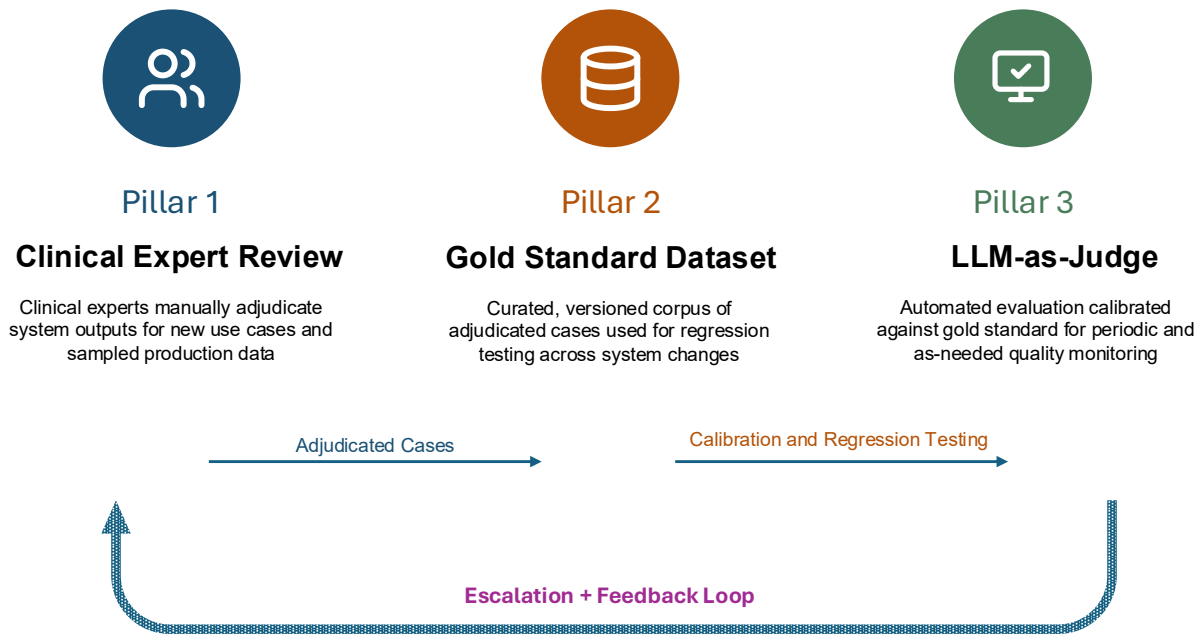

**Figure 8: Evaluation and Monitoring**. In order to validate Scout's ability to handle new use cases, we adopt a framework which combines human verification coupled with ongoing validation against known gold-standard cases. This framework also allows for any changes to Scout (*e.g.* model changes, prompt changes) to be quickly validated.

## Discussion

In this randomized, blinded crossover trial across a variety of clinical specialties and tasks, Scout reduced the time required to complete tasks involving searching and synthesizing EHR data by nearly 40%. Participants also reported significantly lower perceived mental demand and effort. Critically, these gains did not sacrifice the quality of the task outputs. Evaluation of the outputs showed that outputs from Scout met prespecified non-inferiority criteria for accuracy, completeness, and relevance, with evidence that Scout outputs were superior in how complete they were. These findings provide early trial-based evidence that LLM-powered EHR tools can meaningfully reduce both the mental and time burden associated with clinical and administrative tasks without compromising quality.

Notably, this 40% reduction may represent a conservative estimate of the tool's real-world impact. The structured tasks developed for the trial were designed to be completed in a reasonable timeframe of about 5-15 minutes. In practice, many of the tasks that we observed during our pilot deployment such as populating large registry forms or determining clinical trial eligibility are substantially more time-consuming and benefitted immensely from Scout. As the complexity and volume of information increases, the relative advantage of AI-assisted search and synthesis over manual curation is likely to increase.

An interesting finding in the NASA-TLX component subscales was that, despite the quality being judged as non-inferior (and in the case of completeness, superior), participants rated their perceived performance on the tasks as slightly worse than when using only the EHR to complete the tasks. Several mechanisms may contribute to this finding. When participants manually curate information from the EHR, the act of locating and curating the data may provide a sense of familiarity with the source material, which may increase confidence relative to reviewing AI-generated output. Even with Scout's citations and evidence cards, reviewing this output requires trust in a system whose internal reasoning is not as transparent. Additionally, participants had only brief training with Scout prior to the trial, and it is plausible that this performance perception gap would narrow with increased familiarity and experience.

Scout and similar tools shift the core task from manual curation to verification. Finding fragmented pieces of information scattered across multiple note types and data sources presents an enormous burden to clinical staff. This is exactly the work that AI-assisted tools are designed to perform. As a core design principle, Scout provides transparency and provenance for every claim through highlighted spans of text and evidence cards, making it easier both to verify the output from Scout and to help ground the outputs to reduce the number of hallucinations and unsupported claims. The NASA-TLX results are consistent with this design philosophy. The largest improvements were observed in mental demand, effort, and temporal demand, which are tied directly to the manual curation of data. However, frustration did not significantly change, likely reflecting the inherent cognitive cost of verification and the adjustment to a different workflow. Similarly, the slight decrease in perceived performance discussed may be a result of this same shift. Users likely feel less ownership over the output and are less certain about its correctness. An important future area of study would be to see whether these findings change with sustained use. Despite somewhat minimal training and having to learn a new tool, this approach still significantly reduced time and effort burden while maintaining overall quality, which we view as a positive direction for the integration of generative AI tooling into clinical workflows.

**Adoption and Real-World Demand**

The pilot deployment provides additional evidence for the practical value of these tools. Over three months, more than 200 users across more than 20 specialties and service lines generated over 6,000 conversations with Scout. Users of scout spanned faculty physicians, trainees, nurses, pharmacists, and clinical quality leaders, indicating that the demand for these tools is not confined by a single role or specialty. The diversity of observed task types also supports the generalizability of Scout's utility across clinical and administrative workflows.

**Related Work**

Tools that provide LLM-based interfaces to EHR data are emerging across multiple health systems and vendors, reflecting the growing recognition of the potential of these tools to drastically reduce the documentation burden. These tools are likely to converge in design as more complex use cases are identified and common design patterns emerge. Scout's distinguishing features include its citation-first design philosophy and its agentic orchestration, which supports multi-step reasoning and complex queries. For example, if a user asks about the effect of a particular medication on a patient's laboratory values, Scout can first identify when medications were ordered and then search for the relevant laboratory values within the relevant time window. These capabilities reflect our view that these tools have the potential to support increasingly complex workflows while making the verification step efficient in the process.

Methodologically, this work offers early trial-based evidence for the efficacy of these tools in reducing administrative burden while maintaining output quality. Since these systems vary in their design and implementation, including which models are used to complete tasks, it is important for each implementation to measure efficacy and safety independently to verify that solutions have the intended benefit.

**Evaluation, Monitoring and Continuous Re-validation**
A key challenge in the design and implementation of these systems is that they are rapidly changing. Large language models and their capabilities are advancing at a breakneck pace, and tooling designed for one model may not be optimal when new models are released. Our approach combines clinician review of outputs for new use cases, comparisons against a growing gold-standard dataset of adjudicated responses, and automated LLM-as-judge checks that can flag potential performance regressions. We view this as an initial, but necessary step towards the continuous evaluation infrastructure that will be required as the tools increasingly become standard components of clinical workflows.

**Limitations and Future Work**
Several limitations should be considered when interpreting these results. This study was conducted at a single academic health system with a small number of participants per service line, and results may not fully generalize to other institutions, EHRs, or practice settings. While the crossover design and multi-specialty scope strengthen internal validity, external validation across sites will be an important next step. Our evaluation scheme featured a single evaluator per use case to ensure domain expertise and familiarity with the expected outputs. However, additional evaluators per use case would further strengthen the assessment. The LLM-as-judge analysis, while directionally consistent with expert evaluations, is sensitive to model choice and prompt structure.

This work opens several promising directions. The gold-standard dataset assembled through this trial and the ongoing monitoring infrastructure provide a foundation for continuous validation that could serve as a model for other health systems deploying similar tools. A critical question that needs to be answered is how trust and perceived performance evolve with sustained use of these tools. In addition, more sophisticated error analysis, including evaluation of errors of omission and commission, is needed to fully characterize the failure modes of these tools and inform areas of improvement. Finally, as these tools mature, longitudinal studies examining whether efficiency gains are sustained over time and whether they translate into better downstream outcomes for providers and patients are essential.

## Conclusion

In a randomized crossover trial across seven clinical specialties, Scout reduced task completion time by nearly 40% and significantly lowered perceived workload without compromising output quality. These findings provide evidence that LLM-powered EHR tools can meaningfully reduce the burden associated with EHR-related tasks.

Scout's design philosophy follows naturally from the nature of the problems causing this burden. Clinical staff need to trust and verify AI-generated output, which requires grounding claims to the original source data. Clinical and administrative workflows are varied and unpredictable, which require the flexibility of free-text input rather than relying on individual point solutions. In addition, the complexity of modern patient records and their synthesis requires agentic systems that are capable of multi-step reasoning. Together, these features aid in reducing the time and cognitive effort required to search for, synthesize, and verify information across a variety of tasks.

Deployment of these systems requires continuous quality monitoring, governance processes for validating the efficacy of such tools within new use cases, and structured training programs to ensure proper usage and understanding of AI-generated outputs. As the capabilities of these systems continues to improve, we believe that this combination of rigorous evaluation and institutional commitment and resources is necessary to ensure that these tools are deployed safely and effectively. Ultimately, by alleviating the burden on clinical staff, these tools can allow clinical staff to dedicate more time and attention to their patients, improving the quality of care delivered.

## Tool Usage

AI tools were used to aid in the editing of this manuscript under human oversight. All scientific content, analysis, and conclusions reflect the authors' original work and judgment. All final content was read and approved by the authors.

# Appendix A: Scout Technical Details

Scout is an agentic AI system consisting of 3 main components:

1. Data Retrieval
2. Orchestration and tool calling
3. Answer creation and evidence card generation

All data retrieval is done via the FHIR[39] protocol. Data elements which are available at the time of writing include all written note types, diagnostic imaging reports, pathology reports, vital signs, laboratory values, ordered medications, vaccination history, demographic information, previous encounter history metadata, and allergy information. More data types and sources are validated and added on a regular basis. Data is pre-fetched when a user requests information about a particular patient in order to reduce latency and lower the effective time to first token.

All of the LLMs used as part of the Scout system are currently from the GPT family of models from OpenAI, available through Microsoft Azure with appropriate Zero Data Retention (ZDR) policies and agreements to allow for Protected Health Information to securely be passed into the system. When a user enters a query, an orchestrator creates a plan for task execution. For simple tasks only requiring information from note data, the system simply calls the note agent, which begins the process of retrieving up to date note information, identifies the spans within relevant notes which may be useful in answering the question, and construct the final answer. For more complicated tasks which involve multiple steps, the orchestrator selects from a series of tools (implemented as subagents) associated with data domains, such as pulling relevant lab values (this spawns a subagent which reads the user's query and determines how to best filter relevant labs). Finally, this data is compiled altogether and the answer is generated. Each tool call and subagent retains links to the unique ID of the data elements that are being accessed and passes those through to the final answer. This enables the generation of evidence cards, which allow the user to view which pieces of information were used in answer generation. All system interactions are logged according with internal policy and guardrails are applied at the answer in order to ensure safety and compliance. Figure A1 shows Scout's user interface, where users can select filter types (time range, encounters), upload documents, and enter their query.

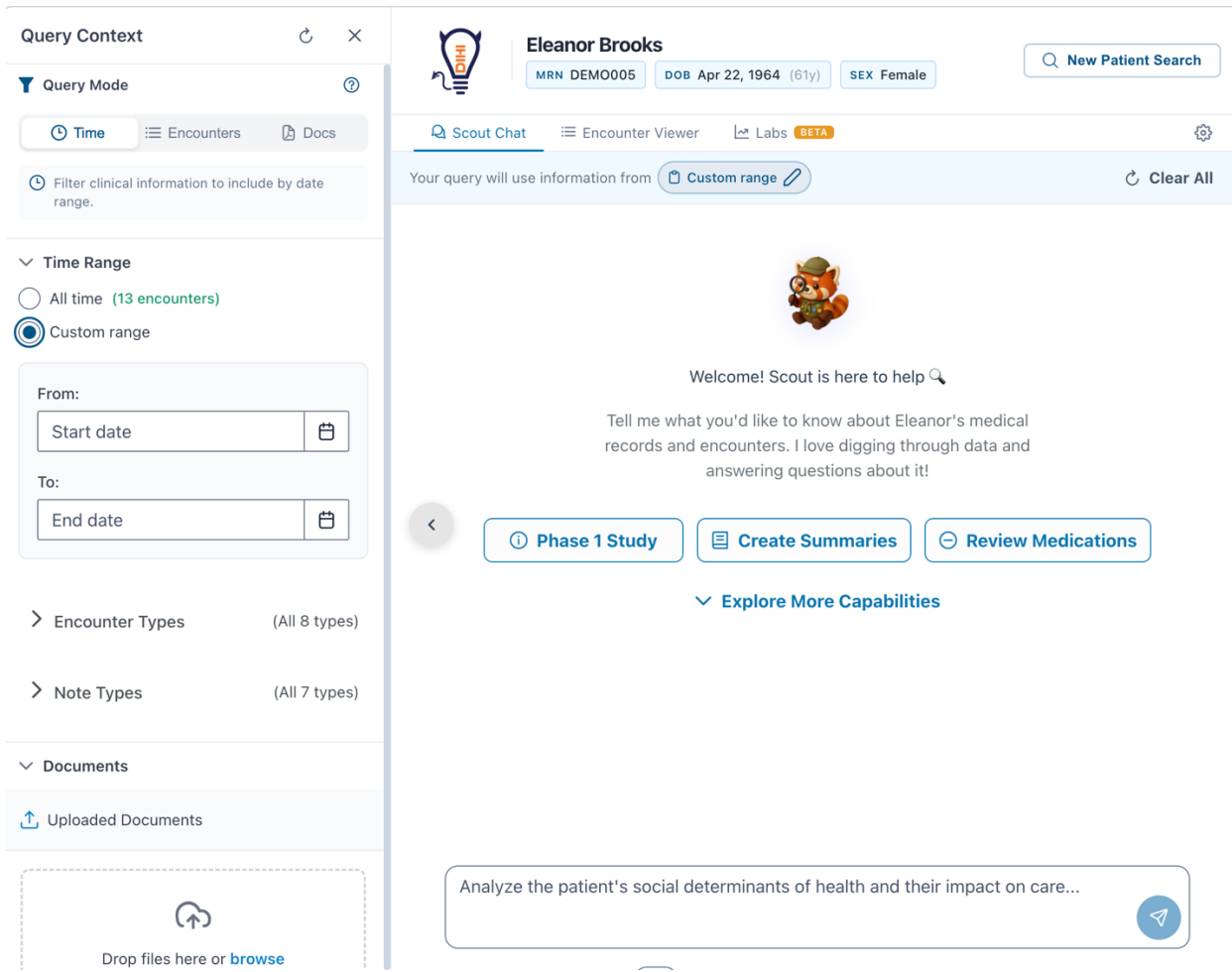

**Figure A1: Scout's User Interface.** The left sidebar allows the user to subset their query by time, encounters, encounter types, and note types. Users can also upload documents for reference. The main panel features example prompts and allows the user to enter their query.

# Appendix B: Use Cases/Task Descriptions

Below are the use cases/tasks that were created by clinical experts. They are listed by specialty.

**Tumor Board (Oncology)**
For the following patients, please fill out the following template in order to prepare for a tumor board discussion

- Date of initial presentation for current oncologic disease
- Tumor type, including histologic subtype and any relevant pathologic grade
- Tumor primary site
- Sites of active disease
- Clinical stage
- Pathologic stage
- Most recent tumor directed therapy, including immunotherapy, cytotoxic chemotherapy,
- Targeted therapy, and radiotherapy

- Prior therapy history, including immunotherapy, cytotoxic chemotherapy, targeted therapy and radiotherapy, with start and end dates
- Previous oncologic surgeries or biopsies
- History of involvement in clinical trials, including the name of the study drugs, where relevant.

## Nephrology

When seeing a patient with chronic kidney disease (CKD) for a follow-up appointment, it is important to capture any recent changes or interval events that have taken place since their last visit.

For your use case, you are to complete the following template:

1. What is the patient's kidney function trend since the last visit? (Compare eGFR/creatinine now vs. prior visit; calculate slope if enough data; highlight new AKI episodes superimposed on CKD.)
2. How has proteinuria/albuminuria changed? (Summarize most recent urine ACR/PCR vs. prior values.)
3. What is the patient's blood pressure trend (home + clinic) since last visit? (Highlight averages, extremes, changes in regimen.)
4. Were there Any cardiovascular events or hospitalizations since the last visit? (MI, CHF exacerbations, stroke, arrhythmias, procedures — linked to kidney status.)
5. Have any changes in medications relevant to CKD have occurred? (New/adjusted: RAAS blockade, SGLT2 inhibitors, diuretics, NSAIDs, nephrotoxins, GLP1ra.)
6. Any medication adherence issues or reported side effects? (Notes on missed doses, cost barriers, hyperkalemia limiting therapy, intolerance.)
7. What is the status of CKD complications since last visit? (Anemia: Hb/iron indices; bone/mineral disorder: Ca, Phos, PTH, vitamin D; metabolic acidosis: bicarbonate; potassium trends. Show as a table.)
8. Are there documented new or worsening symptoms potentially attributable to CKD? (Fatigue, pruritus, edema, anorexia, nausea, cognitive changes, uremic features, foamy urine. Show as a table.)
9. Is there evidence of CKD progression or approaching thresholds for advanced care planning? (eGFR <20, dialysis education referral, transplant evaluation, palliative discussions.)
10. What interval social/functional changes may affect management? (Hospitalizations, new caregiver needs, adherence barriers, diet/transportation issues, insurance/financial stressors.)

## Transplant

Patients who are kidney transplant referrals/candidates who have completed their initial evaluations (in-person evaluations and associated tests/consults) are ready for further discussion during the Kidney Transplant Multidisciplinary meeting. During this meeting, it is important to surface important background information about the patient so that meeting members can be up to date and have a complete picture of the patient's kidney function and test results, etc.

For your use case, you are to complete the following template:

- Describe the cause of CKD/ESRD in 1-2 sentences.
- List the CoMorbid illnesses in bullet form (e.g. History of DM, Symptomatic PVD, Previous Malignancy, Hypertension, Heart Disease)
- Provide the Dialysis start date (MM/DD/YYYY)
- List Sensitization events (pregnancy, blood transfusions or previous transplants)
- Provide the test name and its date (MM/DD/YYYY) for the patient's
  - Last/most-recent echo
  - Last/most-recent stress test
  - Last/most-recent CT scan
- Summarize relevant recent cancer screening (e.g., colonoscopy, Pap smear, mammogram)
- In a few sentences, summarize the notes about
  - social work

- psychology
  - finance
  - nutrition

**Pediatric Cardiology**

Currently cardio-thoracic surgical conferences are used to assess a patient's' readiness for cardio-thoracic surgery. This review features important information about the patient's cardiovascular condition and recent changes.

For your use case, you are to complete the following template:

- Summarize the patient's form of congenital or acquired heart disease in 1-3 sentences.
- Summarize the patient's history of surgical procedures to-date as a bulleted list and in chronological order.
- Summarize the findings of the patient's most recent cardiac catheterization, in bullet points.
- Summarize the findings of the patient's most recent cardiac MRI/MRA and/or chest CT/CTA imaging in bullet points.
- Summarize the findings of patient's most recent electrocardiogram (ECG) in 1 sentence.
- Summarize the findings of patient's most recent ambulatory cardiac monitoring in 1 sentence.
- Summarize the findings of the patient's most recent exercise stress test in 1 sentence.
- Summarize the patient's family history relevant only to cardiovascular disease, as a bulleted list.
- Summarize the patient's cardiac medications and dose adjustments within the last year in chart form.
- Summarize the patient's non-cardiac medical history as a bulleted list.

**Emergency Medicine**

Your use case involves review Emergency Department stays in order to determine more appropriate diagnoses. Oftentimes, the Emergency Department diagnosis will have a generic diagnosis such as "chest pain," even though more information is often made available during the ED that can inform a more detailed and specific set of diagnosis codes. This case is built to mimic this type of review.

For your use case, you are to complete the following template:
*Keep the questions in your report and respond beneath each question, keeping them in order*

1. In one sentence or less, describe the initial visit reason or admitting diagnosis
2. What was the final charted ED clinical impression?
3. Create a bullet list of documented pertinent positive symptoms for this current ED visit related to the clinical impression upon ED-disposition?
4. Create a table with the ED-relevant tests and results.
5. Create a table with the in-ED consults, time of consult, and 1-liner outcome.
6. Create a bullet list of the documented treatments and responses during Emergency care.
7. Create a bullet list of documented post-ED plans for the patient (written in the ED. e.g. admitted for operative management or discharged with Neurology follow up).
8. What text would you update the clinical impression with in order to prepare more-clear and specific phrasing upon ED disposition

**Behavioral Health**

Use Case:

**Objective**: Analyze the patient's electronic medical record to determine whether they meet criteria for treatment-resistant depression (TRD).

**Additional Context**
- Inclusion Criterion: PHQ-9 scores ≥15 across two or more encounters and continues to be ≥15.
- Inclusion Criterion: At least two failed trials of antidepressants from different pharmacologic classes.
- Plan for an Output Format: of a chronological table of medication trials, a timeline of PHQ-9 scores, and a narrative summary with clinical reasoning for TRD diagnosis

**Use Case test to complete (Keep questions in report, use this template):**

1. Extract and organize all psychotropic medication trials into a chronological table. Group medications by class (e.g., SSRI, SNRI, atypical antipsychotic, TCA, MAOI, mood stabilizer) and include augmentation strategies (e.g., lithium, antipsychotics, psychotherapy). Complete a table for this with the with the following columns:
   a. Medication Name
   b. Class
   c. Start Date
   d. End Date
   e. Duration
   f. Dosage
   g. Side Effects
   h. Reason for Discontinuation

| Medication Name | Class | Start Date | End Date | Duration | Dosage | Side Effects | Reason for Discontinuation |
|---|---|---|---|---|---|---|---|
|  |  |  |  |  |  |  |  |

2. **Extract PHQ-9 scores** from the record and present them in a **timeline format**, aligned with medication trials:
   a. Date
   b. PHQ-9 Score
   c. Notes (e.g., linked visit, medication at time)

| Date (MM/DD/YY) | PHQ-9 Score | Notes (e.g., linked visit, medication at time) |
|---|---|---|
|  |  |  |

3. **Assess treatment resistance**:
   a. Confirm ≥2 adequate antidepressant trials from different classes
   b. Evaluate duration and dosage adequacy of each trial
   c. Identify persistent depressive symptoms despite treatment
   d. Note any augmentation strategies used

4. **Generate a clinical summary explaining why the patient may meet TRD criteria, referencing:**
   a. Failed adequate trials
   b. Persistent PHQ-9 scores ≥15
   c. Lack of remission despite multiple interventions

**Adult Cardiology**
**Background:**

In cardiology, a patient can often present with chest pain, but require further examination and preparation in order to better understand the risk for cardiovascular events, further testing, etc. This case is meant to simulate gathering information for a patient with chest pain.

For your use case, you are to complete the following template:

1. **Prior descriptions of pain/discomfort**
* Symptoms related to physical / mental stress?
* Relieved by nitroglycerin (Yes/No)
* Precipitating causes of discomfort
* How long has the discomfort been going on?
* How frequently is the discomfort occurring
* Are there any other associated symptoms (lightheadedness, syncope, pre syncope, nausea, palpitations, etc…)
* Recent cold / cough / uri sx
* Pain reproduced with palpation?
* Recent weight lifting / physical lifting
* General physical activity level (exercise)

2. **Demographics and family history:**
* age
* sex
* Racial / ethnic minority
* Family members with coronary disease (age of family member at diagnosis)
* Family members with sudden cardiac death (age of family member at diagnosis)
* Family members with heart failure (age of family member at diagnosis)
* Family members with any other cardiovascular disease (age of family member at diagnosis)

3. **Medical history**
* Smoking history
* diabetes
* hyperlipidemia
* hypertension
* History of coronary disease
* History of coronary revascularization (stenting / PCI, bypass surgery / CABG)
* History of esophageal reflux disease / esophageal spasm
* Hx of heart failure
* Hx of pericarditis
* anxiety
* fibromylagia
* Depression
* INOCA / cardiac syndrome X

4. **Prior testing and its result (with date relative to current visit date)**
* Total cholesterol
* HDL
* LDL
* Triglycerides
* Lp(a)
* NMR lipoprofile
* echocardiogram
* Chest CT

* Coronary calcium score
* Cardiac MRI
* Any stress testing (there are many types of stress tests)
* EKG(s)
* BNP / pro BNP / NT pro BNP
* CRP / hs crp
* Sedimentation rate

**5. Medications of the following classes:**
* Antianginals (B blockers, CCB, nitrates, ranolazine, ivabradine)
* Antihypertensives
* Anti-lipid agents
* Anti platelet agents
* Anticoagulants
* Diabetic agents (specifically / especially GLP-1, and SGLT2, and metformin)
* Heart Failure GDMT (B blockers, SGLT2, spironolactone, ACE/ARB/ARNI, ivabradine)
* Diuretics (loop and thiazide)

# Appendix C: User Instructions

Example user instructions, which contain instructions on how to complete the tasks described in Appendix A are shown below. These include the randomized patient cases (randomized per use case), along with arm (Scout first vs. EHR-only first).

**Name: \<Participant Name\>**
Case: \<Use Case Name\>

**About:**
You have agreed to participate in the Scout Study, which has been approved by the DUHS Institutional Review Board (IRB), Protocol #00118622. This document contains *specific* information for you to complete as part of the study.

**Instructions:**
Complete the patient cases assigned to you IN ORDER, from 1-10. For *each* patient case, you will track the amount of time it took to complete the case, and will complete a brief post patient case survey. Each use case will be done as follows:

1. Enter the MRN into Scout/Epic depending on whether your patient case is in the WITH SCOUT category or WITHOUT SCOUT category.
2. Begin a stopwatch (we recommend using your phone, though there are online stopwatches available as well).
3. Complete the case instructions to the best of your ability. After it is complete, make sure that it is in the appropriate format (see example below)
4. Once you are ready to submit the case, STOP THE STOPWATCH. Make sure you record the time.
5. Upload your patient case results to the appropriate box folder, with the filename as \<MRN\>_CASE.{docx,pdf,txt,etc.}
6. Complete the post-case survey \<Link to survey\>

\<Use Case Instructions from Appendix A\>

When you feel *comfortable* with the use case, meaning you would be ok with having it reviewed for further action, you may stop the stopwatch. This may mean looking over the information provided by Scout, verifying it with the Evidence Cards/Epic, etc.

**ASSIGNED PATIENT CASE MRNS (To be completed in order)**

**WITH SCOUT**
1  XXXXX
2  XXXXX
3  XXXXX
4  XXXXX
5  XXXXX

**WITHOUT SCOUT**
6.  XXXXX
7.  XXXXX
8.  XXXXX
9.  XXXXX
10.  XXXXX

**Checklist for completing cases:**
1. Enter the MRN into Scout/Epic depending on whether your patient case is in the WITH SCOUT category or WITHOUT SCOUT category.
2. Begin a stopwatch (we recommend using your phone, though there are online stopwatches available as well).
3. Complete the case instructions to the best of your ability. After it is complete, make sure that it is in the appropriate format (see example below)
4. Once you are ready to submit the case, STOP THE STOPWATCH. Make sure you record the time.
5. Upload your patient case results to the appropriate box folder, with the filename as <MRN>_CASE.{docx,pdf,txt,etc.}
6. Complete the post-case survey
   <link to survey>

**Your Personal Upload folder: <Link to upload folder>**
**FAQ**:
1. What happens if I get interrupted?
   a. If possible, pause your stopwatch, then resume it after you are able to get back to the same point that you stopped at earlier. We prefer that you complete each patient use case in one sitting, although we understand that things come up.
   b. If you lose track of the time (stopwatch malfunctions, etc.), please restart the case.
2. Can I use Epic alongside Scout to fill out/verify information during the WITH SCOUT patients?
   a. Yes, you may use both Scout and any additional tools/systems that you normally use. If for instance there is a field that Scout cannot answer and you want to verify it in Epic, you may do so. This time will be counted, and you should only stop the stopwatch when you feel comfortable with your answer
3. I have a question about something related to the study!
   a. Please email us at <coordinator email>
4. What if I am unable to locate specific information for a patient case?

     a. If you cannot find a piece of information in either Scout or Epic, document this in your case notes and proceed with the available data. Do your best to provide a comprehensive answer based on what is accessible.

5. What should I do if I make a mistake or realize a case was submitted incorrectly?
     a. If you notice an error in your case submission, please correct the information and re-upload the file to your personal upload folder. If possible, notify the study coordinator regarding the update.

# Appendix D: NASA-TLX Survey

Thank you for participating in this quantitative study of DIHI's Scout software. This survey asks questions about the task you completed, with or without Scout support. When you used Scout, please evaluate your experience of the task with Scout. This task should take less than a minute to complete. Your responses will be kept confidential.

Please select your netID

[ ⌄ ]

Please provide the Patient's MRN (Duke Qualtrics is secured)

[                                    ]

Time it took to complete the task for the patient

       Minutes                     Seconds

      [      ]                        [      ]

Did you use Scout to complete the task for this patient?

○ Yes

○ No

[ → ]

Please use the sliders to answer each question below

0   5   10   15   20   25   30   35   40   45   50   55   60   65   70   75   80   85   90   95   100

Mental Demand: How mentally demanding was the task?

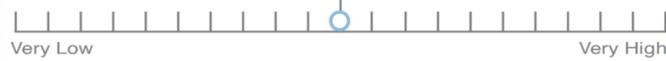
Very Low                                                    Very High

Temporal Demand: How hurried or rushed was the pace of the task?

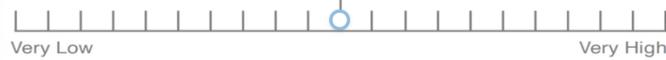
Very Low                                                    Very High

Performance: How successful were you in accomplishing what you were asked to do?

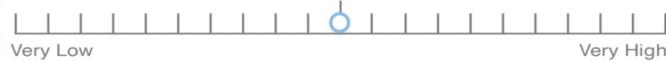
Very Low                                                    Very High

Effort: How hard did you have to work to accomplish your level of performance?

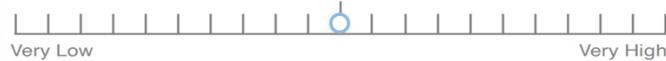
Very Low                                                    Very High

Frustration: How insecure, discouraged, irritated, stressed, and annoyed were you?

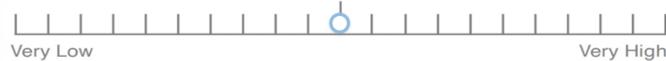
Very Low                                                    Very High

# Appendix E: Quality Adjudication Rubric

Below is the rubric used to adjudicate cases in a blinded manner on the basis of accuracy, completeness, and relevance.

## Accuracy Rubric

| Score | Tier | Description |
|-------|------|-------------|
| 10 | *Exceptional* | Every factual claim is correct and verifiable against the source record. No errors of any kind, including dates, values, medication names/doses, staging, and clinical terminology. Nuance and clinical context are preserved without distortion. |
| 9 | *Excellent* | All clinically material facts are correct. At most one trivial imprecision (e.g., rounding a lab value, minor date approximation) that would not affect clinical reasoning or decision-making. |

| 8 | *Very Good* | No clinically significant errors. One or two minor inaccuracies are present (e.g., a medication dose listed as the prior rather than current dose, a slight misattribution of a finding to the wrong encounter) that would be caught on routine review and would not alter management. |
|---|---|---|
| 7 | *Good* | Generally accurate with most key facts correct. Contains one error of moderate clinical relevance (e.g., incorrect staging modifier, misidentified laterality, wrong medication class) or several minor errors that collectively warrant careful verification. |
| 6 | *Adequate* | The majority of content is accurate but includes two or more moderately significant errors or one error that could plausibly affect downstream clinical decisions if not caught (e.g., incorrect allergy status, wrong procedural date affecting treatment timeline). |
| 5 | *Borderline* | Approximately half of the critical claims are accurate. Multiple errors affecting clinical interpretation are present. The output could still serve as a rough starting point but requires substantial verification and correction before use. |
| 4 | *Below Standard* | Frequent factual errors across multiple domains (medications, labs, history). Some fabricated or hallucinated information is present that has no basis in the source record. Reliability is insufficient for clinical use without near-complete re-verification. |
| 3 | *Poor* | More claims are inaccurate or unsupported than accurate. Contains clearly hallucinated content (events, diagnoses, or results not present in the record). The output is misleading and could cause harm if used without independent chart review. |
| 2 | *Very Poor* | Pervasive inaccuracies throughout. The output bears only superficial resemblance to the actual patient record. Hallucinated content is extensive. The document cannot be used as a basis for any clinical activity. |
| 1 | *Unacceptable* | Factual content is almost entirely incorrect or fabricated. The output describes a clinical scenario that does not correspond to the patient in question. Use of this output in any clinical context would pose a direct safety risk. |
| 0 | *No Output* | No response was generated, or the response contains no factual content relevant to the patient (e.g., refusal, off-topic text, system error). |

## Completeness

| 10 | *Exceptional* | Every element specified in the task rubric is fully addressed. All clinically relevant details are included with appropriate depth. No gaps, omissions, or areas requiring follow-up. The output is ready for its intended clinical or administrative purpose without supplementation. |
|---|---|---|

| 9 | *Excellent* | All major task elements are addressed comprehensively. At most one minor detail is absent (e.g., a single historical date, a non-critical lab value) that does not affect the overall utility or clinical interpretability of the output. |
|---|---|---|
| 8 | *Very Good* | Nearly all required elements are present and adequately detailed. One or two secondary items are missing or underdeveloped (e.g., a follow-up plan mentioned but not elaborated, a screening test omitted from a list), but the core clinical picture is intact. |
| 7 | *Good* | Most task elements are addressed. One moderately important element is missing or insufficiently detailed (e.g., prior treatment history is summarized but missing dates or durations, a required consult summary is absent), requiring supplementation for the intended use. |
| 6 | *Adequate* | The output covers the primary clinical question but has notable gaps in two or more task-specified domains. The information provided is usable but would require meaningful supplementation to meet the task objectives fully. |
| 5 | *Borderline* | Roughly half of the required task elements are addressed. Key sections are either missing entirely or present only in superficial form. The output provides some useful information but is insufficient as a standalone work product. |
| 4 | *Below Standard* | Fewer than half of the required elements are adequately addressed. Major components of the task (e.g., an entire domain such as medication history or imaging summary) are absent. The output requires extensive supplementation. |
| 3 | *Poor* | Only a few elements of the task are addressed, and those that are present lack necessary detail. The output captures fragments of the required information but is not usable for the intended purpose without near-complete reconstruction. |
| 2 | *Very Poor* | Minimal content is provided. One or two elements are touched upon superficially while the vast majority of the task requirements are unmet. The output provides negligible value toward completing the task. |
| 1 | *Unacceptable* | The response contains almost no relevant content. The task requirements are essentially unaddressed. The output is functionally equivalent to a non-response for the purposes of the intended clinical or administrative workflow. |
| 0 | *No Output* | No response was generated, or the response is entirely off-topic, containing no content related to the assigned task. |

## Relevance

| Score | Tier | Description |
|---|---|---|
| 10 | *Exceptional* | Every piece of information included is directly pertinent to the task and clinical context. The output is precisely scoped to the question asked, with no extraneous content. Prioritization of information reflects sound clinical judgment appropriate to the use case. |
| 9 | *Excellent* | Virtually all content is directly relevant. At most one minor tangential detail is included (e.g., a historical finding with marginal bearing on the current question) that does not distract from the core response or reduce usability. |
| 8 | *Very Good* | The output is predominantly relevant and well-targeted. A small amount of peripheral content is present (e.g., a medication listed that is unrelated to the clinical question, an extra historical detail) but does not meaningfully impair readability or efficiency of review. |
| 7 | *Good* | Most content is relevant, but the output includes some clearly extraneous material (e.g., an unrelated systems review, irrelevant social history details) or organizes information in a way that obscures the most pertinent findings. Minor reframing would improve utility. |
| 6 | *Adequate* | The core relevant information is present but is diluted by notable amounts of tangential or off-topic content. The reviewer must actively filter useful information from noise. Alternatively, relevant content is poorly prioritized, burying critical details. |
| 5 | *Borderline* | Approximately equal parts relevant and irrelevant content. The output addresses the general clinical domain but frequently deviates from the specific task or includes substantial information that does not serve the stated purpose. |
| 4 | *Below Standard* | More content is irrelevant than relevant. The output may address a related but distinct clinical question or include extensive boilerplate that does not pertain to the patient or task. Extracting useful information requires considerable effort. |
| 3 | *Poor* | The output is largely off-target. Only isolated fragments address the actual task. The majority of content pertains to unrelated clinical domains, generic templates, or information that does not apply to the patient in question. |
| 2 | *Very Poor* | Nearly all content is irrelevant. The output may superficially reference the correct patient but addresses an entirely different clinical question or workflow. It provides almost no actionable information for the assigned task. |
| 1 | *Unacceptable* | The response is entirely off-topic or addresses a different patient, a different clinical scenario, or a non-clinical subject altogether. No portion of the output is usable for the intended purpose. |

| 0 | *No Output* | No response was generated, or the response contains no substantive content (e.g., system error message, blank output, refusal without clinical content). |

# Appendix F: Sample Size and Power Calculation

Power was evaluated via Monte Carlo simulation to match the planned fixed-case, block crossover study design. Each simulated dataset comprised N participants, each completing the same 10 pre-selected patient cases (MRNs) in a fixed order. Tool assignment followed a two-period block crossover: cases 1–5 were completed with one tool and cases 6–10 with the other. Participants were randomized 1:1 to Scout-first vs Control-first sequences, determining which tool was used in the first block.

**Data-generating model**

Outcomes were generated from a linear mixed model with a participant-specific random intercept and fixed case effects:

$$Y_{ic} = \mu + \beta \cdot \text{Tool}_{ic} + \gamma_c + u_i + \varepsilon_{ic},$$

where $i$ indexes participant and $c$ indexes MRN (1–10). $\text{Tool}_{ic}$ is an indicator for Scout (vs Control). Case effects $\gamma_c$ were treated as fixed, representing systematic difficulty differences among the specific MRNs used in the study; for each scenario, $\gamma_c$ values were drawn once from a mean-zero normal distribution and then held constant across replicates. Participant random intercepts were generated as $u_i \sim N(0, \sigma_u^2)$ and residual errors as $\varepsilon_{ic} \sim N(0, \sigma_e^2)$, independent of each other.

**Analysis model and power definition**

Each simulated dataset was analyzed using the pre-specified primary model:

$$Y \sim \text{Tool} + \text{MRN} + (1 \mid \text{participant}),$$

fit by REML using `lme4::lmer`. Two-sided p-values for the Tool effect were obtained using Satterthwaite's degrees-of-freedom approximation (`lmerTest`). Power was defined as the proportion of simulation replicates with $p < 0.05$ for the Tool coefficient.

**Simulation settings and effect sizes**

For each variance scenario, 5,000 datasets were simulated at N = 20 participants.

**Time endpoint**

The primary time endpoint was analyzed on the log scale. The target effect was a 25% reduction in geometric mean time for Scout vs Control:

$$\beta = \log (0.75) \approx -0.288.$$

Because pilot data were insufficient to estimate variance components reliably, power was evaluated across a grid:

$$\sigma_u \in \{0.20, 0.30, 0.40\}, \sigma_e \in \{0.20, 0.30, 0.35\}$$

(on the log scale). Case-effect heterogeneity was simulated with SD 0.75 (log scale).

**NASA-TLX endpoint**

The targeted effect was an 8-point reduction on the raw NASA-TLX scale (0–100):

$$\beta = -8.$$

Variance components were evaluated across:

$$\sigma_u \in \{5, 8, 12\}, \sigma_e \in \{10, 14, 18\}$$

(in points). Case-effect heterogeneity was simulated with SD 8 points.

Results of simulation grid

For the time endpoint, estimated power was $\geq 0.9998$ across all $(\sigma_u, \sigma_e)$ combinations. For NASA-TLX, estimated power was $\geq 0.8524$ across all combinations; the minimum occurred at $\sigma_u = 8$ and $\sigma_e = 18$. Model-fitting failures were not observed in the simulated scenarios (empirical failure rate 0).

Reproducibility

Seeds were set to ensure reproducibility.

Code for the power simulation can be found at `https://github.com/dihi/scout-preprint-code`

# Appendix G: R Package Version List

```
R version 4.1.3 (2022-03-10)
Platform: aarch64-apple-darwin20 (64-bit)
Running under: macOS 15.1

Matrix products: default
LAPACK: /Library/Frameworks/R.framework/Versions/4.1-
arm64/Resources/lib/libRlapack.dylib

locale:
[1] en_US.UTF-8/en_US.UTF-8/en_US.UTF-8/C/en_US.UTF-
8/en_US.UTF-8

attached base packages:
[1] stats     graphics  grDevices utils     datasets  methods
base

other attached packages:
[1] emmeans_2.0.1  lmerTest_3.2-0 lme4_1.1-37    Matrix_1.4-0
[5] janitor_2.2.1  tidyr_1.3.2    stringr_1.6.0  dplyr_1.2.0

loaded via a namespace (and not attached):
 [1] Rcpp_1.0.8.3       pillar_1.11.1      compiler_4.1.3
 [4] nloptr_2.2.1       reformulas_0.4.1   tools_4.1.3
 [7] boot_1.3-28        digest_0.6.29      gtable_0.3.0
[10] lubridate_1.8.0    evaluate_0.15      lifecycle_1.0.5
[13] tibble_3.3.1       nlme_3.1-155       lattice_0.20-45
[16] pkgconfig_2.0.3    rlang_1.1.7        cli_3.6.5
[19] rstudioapi_0.13    yaml_2.3.5         mvtnorm_1.1-3
[22] xfun_0.30          fastmap_1.1.0      withr_2.5.0
```

```
[25] knitr_1.38          generics_0.1.2       vctrs_0.7.1
[28] grid_4.1.3          tidyselect_1.2.1     snakecase_0.11.1
[31] glue_1.6.2          R6_2.5.1             Rdpack_2.6.4
[34] rmarkdown_2.13      minqa_1.2.5          ggplot2_3.3.6
[37] purrr_1.2.1         magrittr_2.0.3       scales_1.2.1
[40] rbibutils_2.3       htmltools_0.5.2      splines_4.1.3
[43] MASS_7.3-55         rsconnect_0.8.28     xtable_1.8-4
[46] colorspace_2.0-3    numDeriv_2016.8-1.1  utf8_1.2.2
[49] estimability_1.5.1  stringi_1.7.6        munsell_0.5.0
```

# Appendix H: Hallucination Analysis Prompt

We use the hallucination and inaccuracies prompt described in prior work[26] and reproduce them here for posterity.

**Entailment Prompt**

You are an expert clinical NLP adjudicator. Your job is to decide whether the AI-generated text is fully supported by the source text. You will receive: 1. AI-generated clinical content. 2. A set of source chunks from the patient's chart.  Your task: Determine whether every clinically relevant fact stated in the AI-generated content is directly entailed by the source chunks.

Definitions: - A fact is *entailed* if it is explicitly stated or unambiguously supported in the source. - A fact is *not entailed* if it is missing, contradicted, or only inferable by external knowledge. - Ignore stylistic differences. Focus only on factual content.  Output rules (strict): - Return ONLY a JSON object. - The JSON must contain EXACTLY two keys: "all_relevant_facts_entailed" and "explanation". - "all_relevant_facts_entailed" MUST be a JSON boolean: true or false. - "explanation" MUST be a short natural-language rationale (1–2 sentences). - Do NOT include markup, backticks, commentary, or any text outside the JSON. - The output must be valid JSON.  Inputs: <ai_content> {ai_content} </ai_content>  <source_chunks> {source_chunks} </source_chunks>  Expected JSON output format: {{ "all_relevant_facts_entailed": <bool>, "explanation": "<short explanation>" }}

**Classification Prompt**

 You are an expert clinical NLP adjudicator. Your job is to assess the clinical harm posed by inaccuracies or hallucinations in AI-generated clinical text.  You will receive: 1. AI-generated clinical content in <full_ai_output>. 2. A set of explanations identifying facts that are not entailed by the source (i.e., facts not supported by the patient's chart) in <non_entailed_facts>.  Your task: Review the AI-generated content in <full_ai_output> and the provided descriptions of non-entailed facts in <non_entailed_facts>. Then, from those descriptions only, categorize each non-entailed fact as either an inaccuracy or a hallucination, and assess the maximum level of harm the un-edited AI summary could cause if used for clinical care. IMPORTANT: - Do NOT identify new issues. Only categorize the non-entailed facts that are already described in <non_entailed_facts>. - If <non_entailed_facts> is empty or contains no issues, return risk_level: 1, inaccuracies: [], and hallucinations: [].  Definitions: - **Inaccuracy**: Factually incorrect information, contradicting the chart, or misrepresenting certainty (e.g., "probable diagnosis" became "definite"). - **Hallucination**: Made up information that does not correspond to anything existing in the patient's chart.  Based on your review, what is the maximum level of harm the un-edited AI summary could cause if used for clinical care?  Risk Level Definitions: - Level 1 (No harm): The output contains no clinically meaningful errors that would affect patient care.  - Level 2 (Mild harm): Bodily or psychological injury resulting in minimal symptoms or loss of function, or injury limited to additional treatment, monitoring, and/or increased length of stay.  - Level 3 (Moderate harm): Bodily or psychological injury, adversely affecting functional ability or quality of life, but not at the level of severe harm.  - Level 4 (Severe harm): Bodily or psychological injury, including pain or disfigurement, that interferes substantially with functional ability or quality of life.  - Level 5 (Death): The errors could lead to patient death.  Output rules (strict): - Return ONLY a JSON object. - The JSON must contain EXACTLY four keys: "risk_level", "explanation", "inaccuracies", and "hallucinations". - "risk_level" MUST be an integer 1, 2, 3, 4, or 5. - "explanation" MUST be a short natural-language rationale (1-2 sentences) explaining the risk level. - "inaccuracies" MUST be a JSON array of strings, each describing a specific inaccuracy from the non-entailed facts. - "hallucinations" MUST be a JSON array of strings, each describing a specific hallucination from the non-entailed facts. - If no inaccuracies or hallucinations are found in the non-entailed facts, use an empty array []. - Do NOT include markup, backticks, commentary, or any text outside the JSON.

- The output must be valid JSON.  Inputs: <full_ai_output> {full_ai_output} </full_ai_output>  <non_entailed_facts>

```
{expl_no_entail} </non_entailed_facts>  Expected JSON output format:
{{ "risk_level": <integer>, "explanation": "<short explanation>",
"inaccuracies": ["inaccuracy 1", "inaccuracy 2", ...],
"hallucinations": ["hallucination 1", "hallucination 2", ...] }}
```

# Appendix I: Sensitivity Analysis

| Section | Outcome | Tool effect (Scout vs Control) | 95% CI | P value (tool) | P value (Tool×Period) |
|---|---|---|---|---|---|
| Efficiency | Time to completion (ratio of geometric means) | 0.609 | 0.543 to 0.682 | <0.001 | 0.544 |
| Perceived workload (NASA-TLX) | Effort (points) | -17.57 | -22.26 to -12.87 | <0.001 | 0.216 |
| Perceived workload (NASA-TLX) | Frustration (points) | -4.23 | -9.27 to +0.81 | 0.100 | 0.245 |
| Perceived workload (NASA-TLX) | Mental demand (points) | -16.63 | -21.37 to -11.89 | <0.001 | 0.335 |
| Perceived workload (NASA-TLX) | NASA-TLX raw score (points) | -9.26 | -12.74 to -5.78 | <0.001 | 0.406 |
| Perceived workload (NASA-TLX) | Performance (reversed; points) | 3.99 | +0.88 to +7.09 | 0.012 | 0.518 |
| Perceived workload (NASA-TLX) | Temporal demand (points) | -11.47 | -15.61 to -7.34 | <0.001 | 0.370 |
| Evaluator ratings | Accuracy (0-10) | 0.21 | -0.17 to +0.58 | 0.284 | 0.890 |
| Evaluator ratings | Completeness (0-10) | 0.80 | +0.43 to +1.16 | <0.001 | 0.413 |
| Evaluator ratings | Relevance (0-10) | 0.03 | -0.32 to +0.38 | 0.879 | 0.826 |

# Appendix J: Non-inferiority Endpoint Full Results Table

| outcome | Control_mean_ci | Scout_mean_ci | Diff_mean_ci | p_value | margin | noninferior | superior |
|---|---|---|---|---|---|---|---|
| Completeness | 7.39 (6.83, 7.94) | 8.23 (7.67, 8.78) | 0.84 (0.44, 1.24) | 6.06611556702934E-05 | -0.5 | TRUE | TRUE |
| Accuracy | 8.08 (7.69, 8.46) | 8.28 (7.89, 8.66) | 0.20 (-0.20, 0.61) | 0.325060958257031 | -0.5 | TRUE | FALSE |
| Relevance | 8.50 (8.16, 8.84) | 8.55 (8.21, 8.89) | 0.05 (-0.33, 0.43) | 0.784827190964813 | -0.5 | TRUE | FALSE |

# Appendix K: Prompt to Categorize Scout Tasks

EXPERT AI CATEGORIZATION SYSTEM FOR HEALTHCARE EHR QUESTIONS
================================================================
TASK: Categorize 6000+ healthcare questions into a hierarchical system with
strict rules to prevent over-categorization (max 3 categories per question).
Then generate a CSV summary report of the total count of assignments per
category.
INPUT: CSV file with healthcare questions (one question per row)
OUTPUT: XLSX file with 6 columns:
1. Session ID - Copy verbatim from source file
2. Question - Copy verbatim from source file.
3. Best Category - match to the exact category name from the 20 categories
defined below.
4. Interpretation - reasoning for why it matches the assigned category
according to definitions and rules
5. Second Category - When undecided. This could be null
6. Third Category - When very undecided and significant overlap. This could
be null

CATEGORY STRUCTURE (20 categories: 11 primary + 9 subgroups):

PRIMARY CATEGORIES:
1. Clinical Summary and Multi-Chart Review
2. Admissions and Inter-system movement
3. Consults, Internal Handoffs, and Intra-encounter coordination
4. Discharge and Transitions of Care
5. Preventive Care
6. Quality and Safety
7. Clinical Reasoning and Decision Support
8. Clinical Trials and Study-Fit
9. Registries and Forms
10. Point-of-Care Information Retrieval
11. Temporal Pattern Analysis

SUBGROUPS UNDER CLINICAL SUMMARY:
- [Subgroup] Psychiatric Behavioral Health Summary
- [Subgroup] Social and Caregiver Context Summary

SUBGROUPS UNDER POINT-OF-CARE:
- [Subgroup] Point-of-Care Imaging Support and Radiology
- [Subgroup] Point-of-Care Laboratory and Vitals Analysis
- [Subgroup] Point-of-Care Medication Management
- [Subgroup] Point-of-Care Problem Lists and Diagnosis

SUBGROUPS UNDER TEMPORAL PATTERN:
- [Subgroup] Temporal Patterns with Imaging and Radiology
- [Subgroup] Temporal Patterns with Laboratory and Vitals Analysis
- [Subgroup] Temporal Patterns with Medications

STRICT CATEGORIZATION RULES:

Starter clean: If the question is only "Retry", "Rerun", "Do again" then
count these, remove them with the note "USER ERROR", and then provide this as
a caption in the final output.

RULE 1: MAXIMUM 3 CATEGORIES PER QUESTION
- Never assign more than 3 categories total to any question
- If multiple categories match, apply priority rules to keep only top 3

RULE 2: SUBGROUPS REQUIRE PARENT CATEGORY
- If assigning a subgroup, ALWAYS assign its parent category too
- Example: "[Subgroup] Point-of-Care Imaging" requires "Point-of-Care Information Retrieval"
- Example: "[Subgroup] Psychiatric Behavioral Health Summary" requires "Clinical Summary and Chart Review"

RULE 3: MUTUAL EXCLUSIONS (cannot both be assigned)
- Point-of-Care Information Retrieval vs. Clinical Summary (one is quick lookup, one is comprehensive)
- Point-of-Care Information Retrieval vs. Temporal Pattern Analysis (one is current state, one is trends)
- Discharge and Transitions of Care vs. Admissions and System-level movement (different care stage and level emphasis. The former is what happens when the encounter ends; the later is what happens that begins the encounter or cross-encounter, system-level movement over time)
- Discharge and Transitions of Care vs. 'Consults, Internal Handoffs, and Intra-encounter coordination' (One is what happens when the counter ends, one is intra-encounter movement within the same hospitalization or clinic visit).
- Admissions and System-level movement vs. 'Consults, Internal Handoffs, and Intra-encounter coordination' (One is points of health system entry and longitudinal care patterns across encounters, one is unit-level entry and intra-encounter coordination within the encounter)

RULE 4: PRIORITY RULES WHEN MULTIPLE MATCH
- More specific beats more general (Psychiatric Summary > Clinical Summary alone)
- Action-oriented beats information (Discharge > Clinical Summary)
- Temporal beats Point-of-Care if both apply
- Primary intent wins over secondary mentions

RULE 5: SHORT QUESTIONS (<50 chars)
- Require exact keyword matches
- Default most specific category only
- Increase specificity threshold

RULE 6: SPECIAL HANDLING FOR TEMPLATES
- Long questions (300+ chars) with "You are a..." or "Role & Objective" prompts
- Key principle: Templates are verbose but usually have ONE core task. Match the task, not every keyword mentioned.
STEP 1: Identify if it's structured data extraction
- If contains bullets (•), numbering (1. 2. 3.), checkboxes, or requests for multiple (~20+) discrete yes/no answers. If asks to "fill out", "complete", "extract to", or mentions "schema"/"JSON"/"table" with 20+ rows → REGISTRIES AND FORMS
 → then skip remaining template rules
STEP 2: If NOT registries, extract primary clinical intent (ignore specialty-specific details)
- For example, if focus is "discharge" support → Discharge and Transitions of Care only

- For example, if the focus is leading up PRIMARILY to a FUTURE decision to be made about the patient risk (e.g. ESI, mortality, deterioration) → 'Clinical Reasoning and Decision Support'
STEP 3: Add the specialty subgroup ONLY if clear single focus
- If primarily about psychiatric assessment/history (not just mentions psych) → add [Subgroup] Psychiatric Behavioral Health Summary
- If primarily about social determinants/family/support → add [Subgroup] Social and Caregiver Context Summary
- Do NOT add subgroups for other specialties (oncology, genetics, cardiology, etc.) - these stay as Clinical Summary only

INTERPRETATIONS (use these thoughfully to interpret the primary motive of the question):

Clinical Summary and Chart Review: "General questions requesting comprehensive patient summaries, medical histories, review of systems, and clinical documentation for chart review purposes. Includes long template prompts asking for narrative summaries (unless they request structured data extraction for registries/forms)."

[Subgroup] Psychiatric Behavioral Health Summary: "Questions involving psychiatric evaluations, mental health assessments, behavioral health histories, and psychiatric treatment plans (when NOT primarily about medications)"

[Subgroup] Social and Caregiver Context Summary: "Questions about social history, family dynamics, caregiver support, social determinants of health, and living situations"

Admissions and Inter-system movement: "Questions about admission reasons but NOT process (that would be intra-encounter), number of admissions, date of admission, readmissions, and patient movement across different encounters or facilities in the healthcare system. (A facility is a hospital or clinic, NOT a floor or unit)."

Consults, Internal Handoffs, and Intra-encounter coordination: "Questions about coordination and movement WITHIN a single encounter: specialist consults, nurse handoffs, ICU-to-floor transfers within same hospitalization, shift changes, team communication during the same visit or admission. (IF coordination is about what will happen AFTER discharge, then this is not internal but is a 'transition of care')."

Discharge and Transitions of Care: "Questions about discharge planning, discharge summaries, follow-up care, and care transitions to home or another facility (e.g. SNF)."

Preventive Care: "Questions related to health screenings, vaccinations; Advice on smoking cessation, nutrition, weight management."

Quality and Safety: "Questions seeking RETROSPECTIVE information about quality measures, core metrics, patient safety incidents, never-events, mortality patterns, hospital acquired conditions"

Clinical Reasoning and Decision Support: "Questions seeking PROSPECTIVE rationale, evidence, or explanations to support clinical decision-making and differential diagnosis; includes preoperative evaluation, surgical clearance, anesthesia assessment, and perioperative risk evaluation. EXCLUDES reasoning for enrollment in a study or trial - which fits in its own category."

Clinical Trials and Studies: "Questions about clinical trial eligibility, research study enrollment, and experimental treatment protocols"

Registries and Forms: "Requests to complete one or more registries, forms, or VERY LARGE, NON-DISCHARGE-FOCUSED templates. Includes more than twenty questions with bullets/numbering asking for discrete data points, requests to fill out twenty-plus-row schemas/JSON/tables, or more than twenty checkbox-style yes/no questions rather than narrative summaries.

Point-of-Care Information Retrieval: "Quick lookups for specific facts, values, or dates during active patient care - AI as a fast search engine within the EHR"

[Subgroup] Point-of-Care Imaging Support and Radiology: "Questions about specific imaging studies, radiology reports, scan results, and interpretation of diagnostic imaging (CT, MRI, X-ray, PET, ultrasound, etc.)"

[Subgroup] Point-of-Care Laboratory and Vitals Analysis: "Questions about specific lab results, pathology reports, cultures, biopsies, vital signs, and laboratory test interpretations"

[Subgroup] Point-of-Care Medication Management: "Questions about current medications, medication lists, dosing, and what medications patient is taking"

[Subgroup] Point-of-Care Problem Lists and Diagnosis: "Questions about diagnoses, clinical conditions, disease processes, differential diagnoses, and problem list creation"

Temporal Pattern Analysis: "Understanding disease progression, treatment response, and clinical trajectories over time by synthesizing serial measurements and events"

[Subgroup] Temporal Patterns with Imaging and Radiology: "Questions ACROSS imaging studies, radiology reports, scan results showing trends, progression or regression according to image narratives"

[Subgroup] Temporal Patterns with Laboratory and Vitals Analysis: "Questions about lab trends, pathology trends, cultures, biopsy trends, vital signs trends"

[Subgroup] Temporal Patterns with Medications: "Questions about medication changes, medication adjustments, medication contraindications, presented side-effects, medication adherence patterns, and stability on medications over time"

OUTPUT FORMAT:

OUTPUT: json with 4 keys:
1. Question - Summarize or truncate the question.

```
2. Best Category - Exact category name. (Reflects one of the exact category
name from the 20 categories defined above.)
3. Interpretation - reasoning for why it matches the assigned category
according to definitions and rules
4. Second Category - When undecided. This could be null

VALIDATION CHECKS:

After categorization, verify:
1. Calculate average number of distinct categories per question (target: 1.5-
2.0). Average categories per question:    1.5-2.0 (acceptable range)
2. Question distribution: 50-60% get 1 category, 30-35% get 2 categories, 10-
15% get 3 categories
3. No question gets 4+ categories
4. All subgroups have their parent category assigned
5. No mutual exclusion violations

QUALITY INDICATORS:
- 0 subgroups without parents
- <1% of questions with 4+ categories
- Examples clearly demonstrate each category's purpose
- Category counts make clinical sense (Clinical Summary largest, rare
categories <100 questions)
```

# Appendix L: LLM-as-Judge Evaluation Pipeline

**Claim Verification Pipeline**

Scout outputs were first decomposed into smaller text segments representing groups of closely related statements. Each segment was evaluated using an LLM (OpenAI GPT-4.1) acting as a judge model to determine whether the clinically relevant facts contained in the segment were supported by the patient's chart.

Verification proceeded in two stages to account for Scout's citation mechanism and ensure that relevant chart context was considered. First, the judge model evaluated each segment against the specific source spans identified by Scout as supporting evidence for the generated output. If any fact was determined to be unsupported at this stage, the evaluation was repeated using a broader set of source material consisting of the full text of the relevant chart documents. Segments that remained unsupported after this second evaluation step were classified as such.

For segments containing non-entailed facts, a second LLM-based adjudication step categorized the unsupported statements as either a hallucination ("A statement in the generated output that could not be found in or supported by the patient chart") or an inaccuracy ("A statement that contradicted information present in the chart or misrepresented the clinical information documented").

In addition to categorization, the judge model assigned a clinical risk level to the output based on the potential harm that could result if the generated text were used without correction in clinical care. Risk levels ranged from 1 (no harm) to 5 (death), following previously published clinical harm taxonomies.

**Manual Adjudication of LLM Judgments**

To assess the reliability of the automated LLM-as-judge system, we manually reviewed the verification results for a subset of 5 patient cases. For these cases, the automated pipeline decomposed the Scout outputs into 187 discrete claims. The LLM-as-judge system classified 170 claims as supported and 17 as unsupported (9 hallucinations and 8 inaccuracies). Manual adjudication confirmed that all claims labeled as supported were correctly classified. However, 11 of the 17 claims labeled as unsupported were in fact supported by the patient chart, indicating false-positive error by the automated adjudication process.

Several recurring sources of misclassification were observed during manual review. In some cases, the judge model failed to identify supporting evidence present within the chart despite Scout correctly citing the relevant source material. In other instances, the judge model interpreted Scout's synthesized phrasing as incorrect despite the underlying clinical statement being accurate. We also observed cases where the adjudicator model flagged omissions of information that were not required to answer the prompt or were not clinically relevant to the task. Below is an example of an LLM-as-judge explanation that not only failed to locate the supporting information for a claim it deemed unsupported, but also contradicts itself:

```
Most comorbidities listed are supported in problem lists and history,
but not all: PTSD (diagnosed, no current symptoms), chronic
periodontitis/partial loss of teeth, and history of blood transfusions
for anemia/melena are not verifiable in detail from provided sources.
Others, such as chronic periodontitis and tooth loss, are confirmed in
problem lists. Some, like OSA on CPAP, are only partly supported (OSA,
yes; current CPAP use is not confirmed).
```

Among the six claims that were determined to be truly unsupported, two primary error modes were observed: referencing outdated clinical information when more recent documentation contained contradictory findings, and misclassification of symptoms documented in clinical lists. In two additional cases, the generated statements were factually correct but were not linked to the appropriate supporting citations in Scout's output, preventing the automated adjudicator from identifying the relevant evidence in the chart.